\documentclass[twocolumn]{aastex631}

\usepackage{newtxtext,newtxmath}
\usepackage{rotating}
\usepackage[T1]{fontenc}

\DeclareRobustCommand{\VAN}[3]{#2}
\let\VANthebibliography\thebibliography
\def\thebibliography{\DeclareRobustCommand{\VAN}[3]{##3}\VANthebibliography}

\usepackage{graphicx}	
\usepackage{amsmath}	
\usepackage{subfigure}

\usepackage{multirow}

\newcommand{\MHz}{{\rm MHz}}

\newcommand{\rev}[1]{{#1}}
\newcommand{\revnew}[1]{{{#1}}}

\begin{document}
\title{Tackling Challenges in 21cm Global Spectrum Experiment: \\ the Impact of Ionosphere and Beam Distortion}

\correspondingauthor{Xin Wang; Xuelei Chen}
\email{wangxin35@mail.sysu.edu.cn; xuelei@cosmology.bao.ac.cn}
\author{Yue Wang}
\affiliation{School of Physics and Astronomy, Sun Yat-Sen University,  No.2 Daxue Road, Zhuhai 519082, China}
\author{Xin Wang}
\affiliation{School of Physics and Astronomy, Sun Yat-Sen University, No.2 Daxue Road, Zhuhai 519082, China}
\affiliation{CSST Science Center for the Guangdong-Hong Kong-Macau Greater Bay Area, SYSU, China}
\author{Shijie Sun}
\affiliation{National Astronomical Observatories, Chinese Academy of Sciences, 20A Datun Road, Chaoyang District, Beijing 100101, China}
\author{Fengquan Wu}
\affiliation{National Astronomical Observatories, Chinese Academy of Sciences, 20A Datun Road, Chaoyang District, Beijing 100101, China}
\author{Shoudong Luo}
\affiliation{School of Physics and Astronomy, Sun Yat-Sen University,  No.2 Daxue Road, Zhuhai 519082, China}
\affiliation{Zhejiang University, Hangzhou 310058, China}
\author{Xuelei Chen}
\affiliation{Department of Physics, College of Sciences, Northeastern University, Shenyang 110819,  China}
\affiliation{National Astronomical Observatories, Chinese Academy of Sciences, 20A Datun Road, Chaoyang District, Beijing 100101, China}
\affiliation{School of Astronomy and Space Science, University of Chinese Academy of Sciences, Beijing 100049, China}
\affiliation{Center of High Energy Physics, Peking University, Beijing 100871, China}

\date{Accepted XXX. Received YYY; in original form ZZZ}


\begin{abstract}
  The HI 21cm global signal from the Cosmic Dawn and the Epoch of Reionization (EoR) offers critical insights into the evolution of our Universe. Yet, its detection presents significant challenges due to its extremely low signal-to-contamination ratio and complex instrumental systematics.
  In this paper, we examine the effects of the ionosphere and antenna beam on data analysis. The ionosphere, an ionized plasma layer in the Earth's atmosphere,  refracts, absorbs, and emits radio waves in the relevant frequency range. This interaction results in additional spectral distortion of the observed signal, complicating the process of foreground subtraction. Additionally, chromatic variations in the beam can also introduce further contamination into the global spectrum measurement. 
 Notably, the ionospheric effect, being dependent on the direction of incoming light, interacts with the instrumental beam, adding another layer of complexity. To address this, we evaluate three different fitting templates of foreground: the logarithmic polynomial, the physically motivated EDGES template, and a SVD-based template.  Our findings indicate that the EDGES and SVD templates generally surpass logarithmic polynomials in performance.
Recognizing the significance of beam chromaticity, we further investigate specific beam distortion models and their impacts on the signal extraction process.
\end{abstract}




\section{Introduction}

The global 21 cm signal originating from cosmic neutral hydrogen is regarded as one of the most promising tools to study the early structure formation of the Universe. Detecting this signal holds the potential to unveil crucial insights into the timeline and physical process of Cosmic Dawn and the Epoch of Reionization  \rev{\citep{Madau_1997,Pritchard2012,Cohen2018}, including the properties of first stars and galaxies, astrophysical feedback processes \citep{Paolo00,Chen_2004,Furlanetto_2006,Barkana_2018,Mebane2020}, the nature of dark matter \citep{Dvorkin14,Tashiro14,Lopez_2016,Mondal2024}, cosmological parameters and fundamental physics \citep{Tozzi_2000,Sierra2018,Novosyadlyj2023}. }

Several experiments are currently underway or in the planning stages to measure this global signal, including the Experiment to Detect the Global EoR Signature (EDGES, \citealt{EDGES2018Nature}), Shaped Antenna measurement of the background RAdio Spectrum (SARAS,  \rev{\citealt{Girish_SARAS_2020,Raghunathan_SARAS_2021,Nambissan_SARAS_2021}}), Probing Radio Intensity at high-Z from Marion (PRIZM, \citealt{PRIZM17}), The Radio Experiment for the Analysis of Cosmic Hydrogen (REACH, \citealt{de_Lera_Acedo_2022}), The Broadband Instrument for Global Hydrogen Reionisation Signal (BIGHORNS, \citealt{Sokolowski_2015}), Large Aperture Experiment to Explore the Dark Ages (LEDA, \citealt{Price_2018}), \rev{the Dark Ages Radio Explorer (DARE, \citealt{Burns_2017}), Zero-spacing Interferometer \citep{Raghunathan_Zerospacing_2011,mahesh2015resistive} and the Low-frequency Anechoic Chamber Experiment (LACE, \citealt{Huang_2021,liu2023}}).
Primarily due to the faintness of the signal and complex systematics involved, the current status of observations remains a subject of ongoing debate. \citealt{EDGES2018Nature} has claimed a successful detection of the absorption profile, with an amplitude at least twice that of conventional theoretical predictions, \rev{ whereas \citealt{singh2022} has reported a non-detection.} This disparity clearly underscores the challenges and difficulties in such experiment. Therefore, more studies are needed for further understanding.

A major challenge in these experiments is the foreground contamination, which has an amplitude that is four to five orders of magnitude greater than the cosmic signal. The sources of this foreground include galactic synchrotron radiation, free-free emission, dust and extragalactic point sources. 
While dealing with such low signal-to-contamination ratio is not unusual in cosmological observations, it still poses a unique challenge in the 21cm global spectrum measurement. 
In contrast to the low-redshift 21cm intensity mapping, where the cosmological signal is randomly fluctuating across frequencies, the global spectrum signal is much smoother. This complicates efforts to utilize the foreground's smoothness for effective contamination removal. 
\revnew{Numerous studies have undertaken efforts in various aspects of data analysis to overcome this challenge. An overview of foreground removal is provided by \cite{LiuShaw2020review}. 
More specifically, extensive discussions on the impact of foreground modeling \citep{Adrian2013,Nhan2019,Hibbard2020,Anstey2021,Pagano2024} and signal modeling \citep{Harker2016,Shen2024} have been detailed. 
Various methodologies have been proposed, including Markov Chain Monte Carlo (MCMC) \citep{Harker2012,Tauscher2018}, Bayesian methods \citep{Harker2015}, assessments of fit robustness \citep{Tauscher2018_,Tauscher2020_1,Bassett2021}, and several other combined approaches to address the problem  \citep{Adrian2013,Rapetti2020,Tauscher2021,Begin2022,Anstey2023}.
}
Furthermore, in terms of foreground fitting, various fitting methods have been suggested, including logarithmic polynomials \citep{Pritch10,Harker11,Liu12,Hibbard2023}, more physically motivated functional forms \citep{EDGES2018Nature}, and templates specific to particular instruments \citep{Switzer_2014,Ved14}.

The chromatic response of the instrument introduces additional layers of complexity, especially considering the $\sim 10^{-4}$ signal-to-contamination ratio. 
Consequently, any spectral distortion at this level can significantly affect the interpretation of the cosmological signal.
Despite substantial efforts by many experiments to reduce these effects, practical challenges persist.
For example, this includes manufacturing inaccuracies, mechanical structural deformations, and chromatic responses of electronic components, all of which can introduce frequency-dependent fluctuations. 
\rev{Similarly, significant effort has been dedicated to exploring the impact of instrumentation as well, detailed in studies by \citep{Tauscher2018,Hibbard2020,Tauscher2020_2,Tauscher2021,Anstey2021,Anstey2022}. }

Furthermore, in ground-based experiments, the impact of the ionosphere cannot be overlooked either \rev{ \citep{Ved14,Datta2016,Shen21,Shen2022}.} The ionosphere, consisting of an ionized plasma layer in Earth's atmosphere and extending from about $50$ to $600~ {\rm km}$ above the sea level, plays a significant role in the propagation of electromagnetic waves. Within our target frequency range, namely from $30$ to approximately $100 \MHz$, these waves are reflected, refracted and absorbed by the ionospheric plasma. Crucially, these effects exhibits variability across frequency, as well as in spatial and temporal domains, introducing further complexities into the data analysis process.


This paper delves into the challenges faced by the 21cm global spectrum experiment. 
Typically, in such an experiment the instrument is designed to measure the redshifted 21cm global spectrum in the frequency range of $30 ~\MHz$ to $100 ~\MHz$, with minimal chromatic beam variation. 
Our main objective in this paper is to assess the effectiveness of different foreground cleaning techniques with various observational complexities, including the ionospheric effects and beam deformation. This analysis is crucial for evaluating potential enhancements for the experiment's subsequent space-based phase.

This paper is organized as the follows: Section 2 provides an overview of our mock observation process, including aspects such as the instrument design, beam modeling, instrumental noise, the foreground sky model, and the modeling of the global 21cm signal. Section 3 delves into the physics and specific parameters of the ionospheric effect  utilized in this study. In Section 4, \rev{we present a detailed examination of the three foreground templates used in this analysis}, namely the logarithmic polynomial, the EDGES template, and the instrument-specific SVD template. Section 5 investigates the potential impact of beam distortions on the efficacy of these template fitting methods. Finally, in Section 6, we summarize and discuss our key findings. 

\begin{figure*}
    \includegraphics[width=0.9\textwidth]{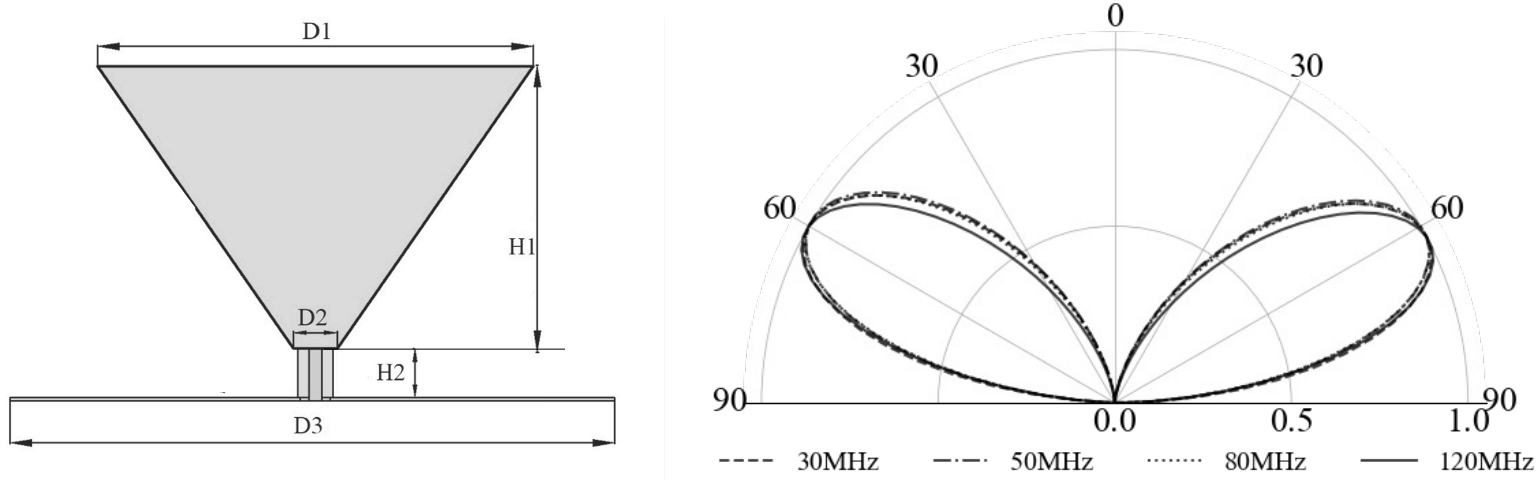}
    \caption{One of the design configurations for our 21cm Global Spectrum Experiment features an inverted cone antenna. As illustrated in the {\it right} panel,this design demonstrates minimal chromatic variation in the normalized beam pattern.   }
    \label{fig:Origin Beam}
\end{figure*}

\section{Mock Observation}
In this section, we begin by providing an overview of our mock observation simulation.  This pipeline, based on our specific beam and sky model, integrates ionosphere effects and instrumental noise to generate time-ordered mock observational data. For concreteness, we assume a particular design of the instrument, which is located at the Tianlai observation site in Hami, Xinjiang, China, at coordinates $90^\circ 48' 24'' E$, and $44^\circ 9'10'' N$, with an elevation of $1500$ meters. 
\rev{We'd like to emphasize that although our baseline calculation is based on a specific beam design and particular geographic location for ionospheric modeling, the conclusions and results derived is broadly applicable.}
Given the significant role and complexity of the ionosphere effect, we provide its detailed  modeling in the next Section.

\subsection{Instrument Design and Beam Pattern}
\label{sec:inst_beam}

The design of the antenna system is guided by several key principals. Firstly, the design focuses on minimizing the chromaticity of the resulting beam. \rev{Secondly, a wider beam width is favored to increase the efficiency of global spectrum observation. }
Thirdly, the return loss of the antenna is designed to be smooth across frequencies, which helps to prevent the introduction of complex functional forms that might complicate the process of foreground fitting. 
Finally, in preparation for upcoming space-borne experiment, the design emphasizes lightweight and compact construction to ensure compatibility with the limited compartment space.  Additionally, it also incorporates features such as high structural strength and vibration resistance, and is engineered to withstand drastic temperature changes.

The design of the antenna system is depicted in the left panel of Figure \ref{fig:Origin Beam}. This single-cone antenna design ensures an almost frequency-independent beam and maintains a stable input impedance across a broad frequency range. 
This is evident from the normalized beam pattern obtained from electromagnetic simulations, as displayed in the right panel for selected frequencies. In the following, we will employ this theoretical model as our baseline beam. However, the real-world beam may deviate from such idealized model, and we will explore some potential deviations and assess their impact on the signal extraction process in Section \ref{sec:beamdistortion}.

\subsection{Instrumental Noise}

In addition to the beam pattern, accurately modeling instrumental error is also essential for our mock observations. Besides the random thermal noise, there are potential systematic errors, \rev{such as those arising from inaccurate measurement of antennas, receivers and other components.}
However, for the scope of this study, our focus will be primarily on thermal noise.
We model the thermal noise as a Gaussian random process with zero mean and \rev{frequency dependent variance} $\sigma^2$, expressed as
\begin{eqnarray}
\sigma = \frac{T_{\rm sys}}{\sqrt{\Delta t \Delta \nu}}, 
\end{eqnarray}
where $\Delta t$ represents the integration time, and $\Delta \nu$ denotes the frequency bandwidth. In our analysis, we have chosen an integration time of $7$ nights and a frequency bandwidth of  $0.2 \MHz$. 
The system temperature, denoted as $T_{\rm sys}$, accounts for the combination of receiver and sky temperature. In this study, we assume the receiver noise temperature to be approximately $100 \mathrm{K}$.

\subsection{Sky Model}

Considering that the foreground emissions are several orders of magnitude stronger than cosmological signals \citep{Fur06}, accurately modeling the foreground is crucial in our end-to-end mock simulation.  Within the frequency range of $30-120 ~\MHz$, the dominant foreground sources include diffuse synchrotron radiation from the Galaxy, free-free emission, and point sources.

In this study, we utilize the Global Sky Model 2008 (GSM2008) \citep{Oli08} to describe the foreground sky, i.e. $T_{\rm FG} = T_{\rm GSM}$. This model provides a whole sky map across large frequency range, developed through the \rev{Principal} component analysis (PCA) of $11$ existing observational maps ranging from $10 \MHz$ to $94 {\rm GHz}$. As detailed in \citep{Oli08}, the dominant contribution below a few ${\rm GHz}$ is the synchrotron radiation. It is worth noting that, in this paper, we refrained from utilizing the updated version of the global sky model \citep{GSM2017MNRAS} due to some non-physical values present in the map, which could potentially affect the mock observation data.

Regarding the cosmological 21cm signal, here we assume a homogeneous distribution of neutral hydrogen on large scales, and a spatially constant 21cm signal in our mock observation. The inhomogeneity may slightly enhance the detected global signal \rev{\citep{Xu_2018,Xu__2021,Munoz2021}}, 
but we neglect this effect here. We further simplify the model by excluding subtle absorption troughs and peaks, and only focus on the absorption profile during the Cosmic Dawn and EoR, as reported by the EDGES experiment. 

We consider two specific signal models. The first is a flattened Gaussian signal, similar to the result reported by EDGES experiment \citep{EDGES2018Nature}:
\begin{eqnarray}
\label{eqn:sig_flatgauss}
T^{\rm flat-Gauss}_{21}(\nu)=-A \left(\frac{1-e^{-\tau e^B}}{1-e^{-\tau}}\right),
\end{eqnarray}
where parameter $B$ takes the form of
\begin{eqnarray}
B=\frac{4(\nu-\nu_0)^2}{\omega^2}\log\left[-\frac{1}{\tau}\log\left(\frac{1+e^{-\tau}}{2}\right)\right]
\end{eqnarray}
Here, $A$ is the amplitude of absorption profile, $\omega$ is the full-width at half-maximum and $\tau$ is the flattening factor. As discussed in \cite{EDGES2018Nature}, \rev{this is a useful functional form to capture the basic shape of the profile.} 
In our study, we adopt parameters that are the same to those reported by \citep{EDGES2018Nature}, an amplitude of $A = 0.53 {\rm K}$, a center frequency of $\nu_0=78.1 \MHz$ and a width of $\omega = 18.7 \MHz$. 
Meanwhile, we also consider the widely used Gaussian model to describe the 21cm absorption absorption profile during:
\begin{eqnarray}
\label{eqn:sig_gauss}
T^{\rm Gauss}_{21}(\nu)=A_0 \exp\ \Bigg({\frac{-{(\nu-\mu)}^2}{2{\sigma}^2}}\Bigg).
\end{eqnarray}
with an amplitude of $A_0 = -0.5 {\rm K}$, at center frequency of $\mu=78.1 \MHz$ and a width of $\sigma = 10 \MHz$. 
To \rev{summarize}, our sky temperature model, denoted as $T_{\rm sky}$, is a composite of the monopole 21cm signal $T_{21}$ and foregrounds emission, 
\begin{eqnarray}
T_{\rm sky}(\nu,\theta,\phi)=T_{21}(\nu)+T_{\rm FG}(\nu,\theta,\phi).
\end{eqnarray}

\section{IONOSPHERE OF EARTH}

Low frequency electromagnetic waves are refracted, reflected, and absorbed by the Earth's ionosphere. A primary focus of this paper is the incorporation of a detailed ionospheric model into our mock simulation pipeline, and examine the spectral distortion caused by this effect. 
In the standard model, the ionosphere is divided into several layers based on the altitude and ionization density, from D, E and F layer. 


The F-layer, also known as the Appleton-Hartree layer, is the uppermost layer of the Earth's ionosphere. Situated at altitudes ranging from approximately 150 to 500 km above the ground, it is primarily composed of atomic oxygen and molecular nitrogen. Due to its high electron density, the F-layer refracts radio waves, acting much like a spherical lens that bends these waves towards the zenith \citep{Ved14}. Consequently, when radio waves are refracted by the F-layer, a ground-based radio antenna can receive signals from a broader section of the sky, resulting in an elevated antenna temperature. 

During the daytime, the F-layer could be divided into two sub-layers, the F1 and F2 layers. The F1 layer is located at an altitude of about 150 to 220 km, while the F2 layer at 220 to 500 km. Although these two layers have different ionization mechanisms and exhibit different diurnal and seasonal variations, modeling them separately does not significantly improve accuracy but adds significant complexity to the simulations. We follow a common simplification used in ionospheric modeling and radio wave propagation simulations that suppose F-layer as a single layer between 200 and 400 km altitude, for the majority of the ionization in the F-layer is concentrated within this range, and at higher altitudes, the electron density decreases rapidly.

The electron density of the F-layer is also affected by solar radiation and the Earth's magnetic field and so on. During the daytime, solar radiation ionizes the atoms and molecules in the F-layer, leads to significant variations in the electron density throughout the day. At night, the absence of solar radiation causes the ionosphere to enter a state of relative stability, and ignoring geomagnetic activities and special spatial weather events, the electron density is primarily controlled by the balance between recombination and diffusion processes, thus we take F-layer as a static model.


The D-layer is the ionosphere's lowest region, positioned at an altitude of approximately $60-90 ~\mathrm{km}$ above the Earth's surface. The high ionization density in the D-layer results from the absorption of both high-energy ultraviolet and X-ray radiation from the Sun, as well as the ionization of nitric oxide (NO) by Lyman-alpha radiation with a wavelength of $121.6\ {\rm nm}$. Moreover, $\mathrm{N}_2$ and $\mathrm{O}_2$ can also become ionized in the D-layer by the hard X-rays ( wavelength shorter than $1\ {\rm nm}$) generated by solar flares.

During the night, the electron density in the D-layer of the ionosphere drops to its minimum level and remains relatively stable, due to the absence of ionizing solar radiation. While it is true that various factors, such as temperature, can still cause fluctuations in the electron density during the night, \rev{in the present model we consider a simple variation.}

At lower frequencies, radio waves interact with the free electrons in the ionosphere causing them to vibrate and collide with other molecules. This interaction leads to the absorption or scattering of radio waves, resulting in a reduction in the strength of the radio signal received by ground-based radio telescopes.

\begin{figure}
    \centering
    \includegraphics[width=8.5cm]{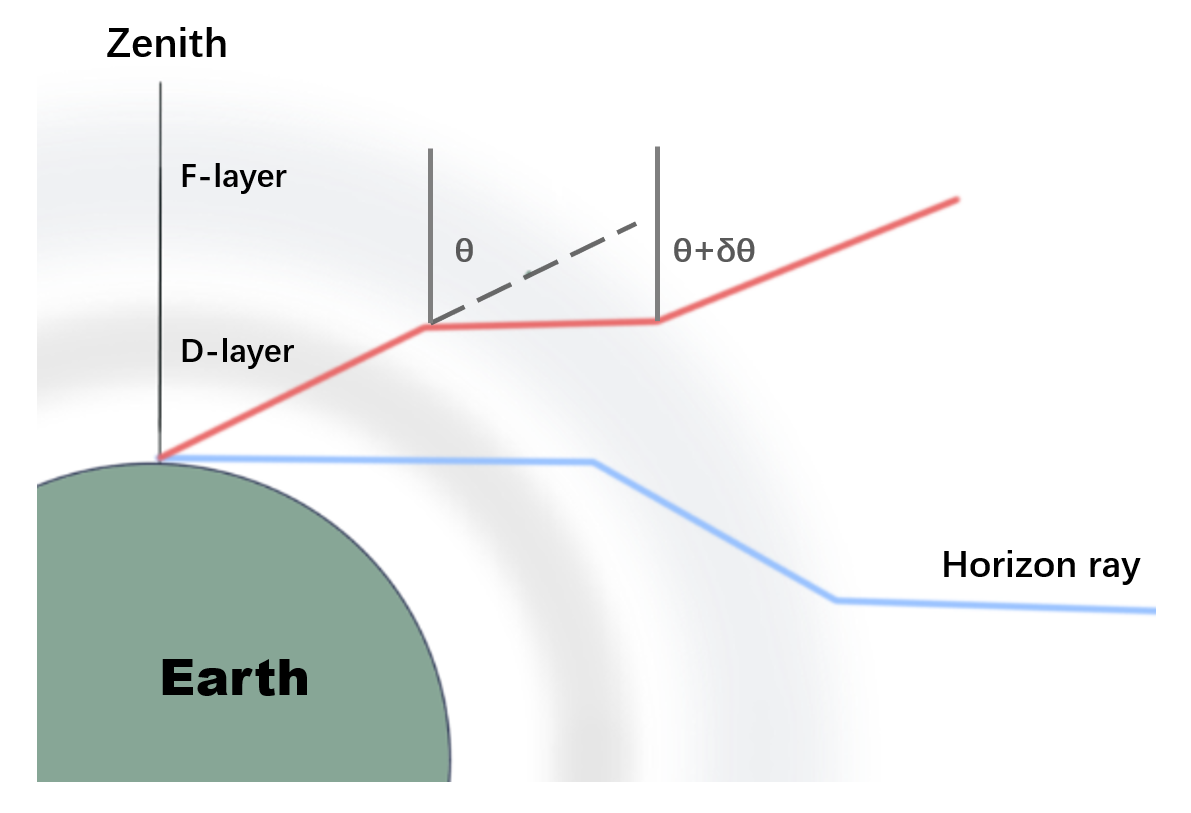}
    \caption{Not to scale simplified representation of ionospheric refraction. The curvature of the Earth causes sources in the sky to appear deviated from their actual positions. In this context, a homogeneous ionosphere functions as a lens, further influencing the paths of incoming signals \citep{Ved14}.}
    \label{fig:refraction}
\end{figure}

\begin{figure*}
    \centering
    \subfigure{
    \includegraphics[width=17cm]{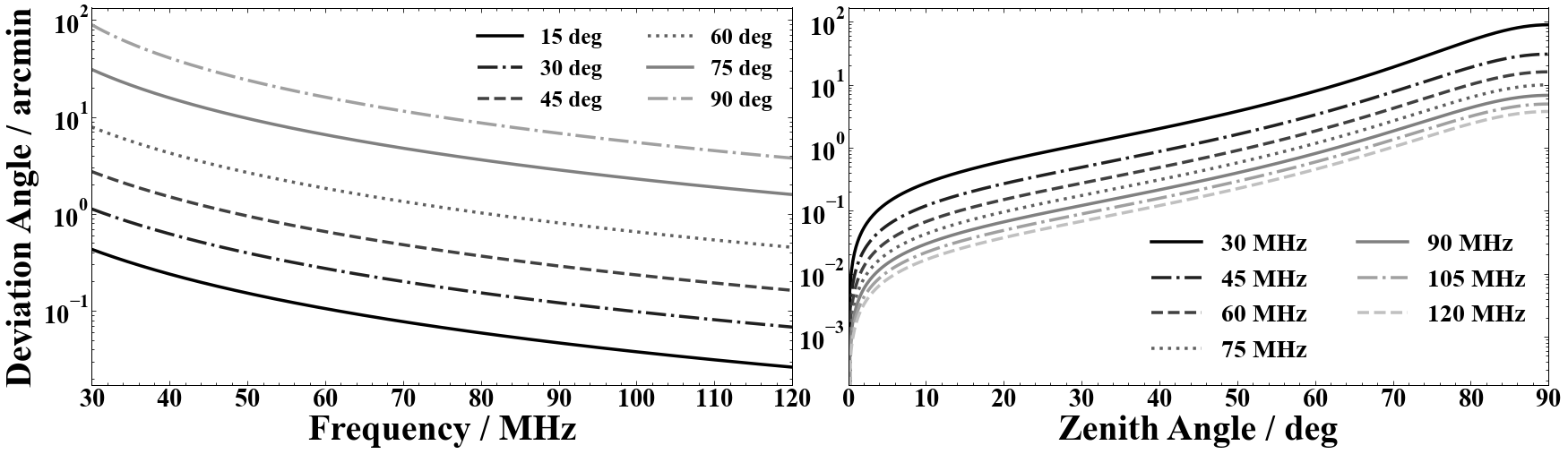}}
    \caption{The refractive effect of a single parabolic F-layer height from $200\sim 400$ km, adopting the model parameters of \citet{Ved14}. Maximum electron density $N_0$ is set to $5\times 10^{11}{\rm m}^{-3}$ at a height $h_m=300$ km. (a) The left figure illustrates the variation of refractive effect caused by the F-layer ionosphere as a function of frequency. The refraction angle becomes zero at the zenith; (b) The right figure illustrates the variation of refractive effect as a function of zenith angle. As the angle of incidence approaches the horizon, the refractive effect becomes stronger, resulting in greater beam distortion.}
    \label{fig:deviation angle}
\end{figure*}

\subsection{Refraction}
The Appleton-Hartree equation describes the refractive index of electromagnetic waves propagating in a cold plasma under the influence of electric and magnetic fields \citet{AH_Equation1961}:
\begin{align}
\eta^2=1-\frac{X}{1-i Z- \left( \mathcal{A} \pm 
\mathcal{B} \right) / ( 1-X-i Z ) }
\label{equation: Appleton-Hartree equation}
\end{align}
where $X={\nu_p}^2/\nu^2$ is the squared ratio of the plasma frequency $\nu_p$ to the electromagnetic wave frequency $\nu$, $Z=\nu_c/\nu$ 
accounts for collisional damping in the plasma, $\nu_c$ denotes the electron collision frequency. The  two angle related terms are $\mathcal{A}= (Y^2 \sin^2 \theta)/2$ and $\mathcal{B}= \left[ \left( Y^4 \sin ^4 \theta\right) /4 +Y^2 \cos ^2 \theta(1-X-i Z)^2 \right ]^\frac{1}{2}$, with $\theta$ being the angle between the ambient magnetic field vector and the wave vector. 
Here $Y=\nu_H/\nu$, which is the ratio of the electron gyrofrequency $\nu_H$ to the wave frequency, which measures the influence of the magnetic field.

Assuming an unpolarized sky over scales comparable to the antenna beam, we have omitted the consideration of Faraday rotation resulting from the Earth's magnetic field, denoted as the $Y$-related terms. 
This simplifies the Appleton–Hartree equation \citep{Ved14,Shen21} to the following form
\begin{eqnarray}
\eta^2 = 1-\frac{X}{1-iZ} = 1-\frac{({\nu_p}/{\nu})^2}{1-i({\nu_c}/{\nu})}
\label{equation: simplified Appleton–Hartree equation}
\end{eqnarray}
The real part of the equation is primarily influenced by the electron density via the plasma frequency $\nu_p$ and induces a deviation in the phase velocity, leading to the refractive effect. 
For a given electron number density $N$, the plasma frequency $\nu_p$ is expressed as:
\begin{eqnarray}
\nu_p = \frac{1}{2\pi}\sqrt{\frac{Ne^2}{m_e\epsilon_0}},
\label{equation: electron plasma frequency}
\end{eqnarray}
where $\epsilon_0$ is the vacuum permittivity.

Due to the relatively high electron density and low atmospheric gas density in the F-layer, it suffices to consider only the real part of Equation (\ref{equation: simplified Appleton–Hartree equation}), so that
\begin{eqnarray}
    \eta_F = \Bigg( 1-\frac{(\nu_p)^2}{\nu^2}\Bigg)^{\frac{1}{2}} ,
\label{equation: real part of sAH}
\end{eqnarray}
To proceed, it is necessary to determine the variation of the electron number density within the F-layer. We assume that the density  can be described by a parabolic model as a function of height \citep{Bailey48,Shen21}.

\begin{eqnarray}
N=N_0\Bigg[1-{\Bigg(\frac{h-h_m}{d}\Bigg)}^2\Bigg],
\end{eqnarray}
where $h$ is the height from earth and $d$ is the semi-thickness of the layer, $N_0$ the maximum electron density within the F-layer at a height $h_m$.
Introducing the critical frequency $\nu_0$ at the vertical incidence of this layer, the refraction index can be further expressed as 
\begin{align}
\eta_F = {\left[1-\frac{N}{N_0}{\left(\frac{\nu_0}{\nu}\right)}^2\right]}^{\frac{1}{2}} 
= 1-{\left(\frac{\nu_0}{\nu}\right)}^2\left[1-{\left(\frac{h-h_m}{d}\right)}^2\right]
\label{equation:F_refraction_index}
\end{align}

Therefore, the deviation angle $\delta \theta$ caused by ionospheric refraction at frequency $\nu$ and angle $\theta$ could be written as:
\begin{eqnarray}
\delta \theta (\nu,\theta) &= &\left(\frac{\nu_0}{\nu}\right)^2\frac{a \cos{\theta}}{d^2} \nonumber \\
&&\times \displaystyle\int_{h_m+d}^{h_m-d} \frac{(h-h_m) \,dh }{\eta_F^2 \left[ (h+a)^2\eta_F^2-a^2\cos^2{\theta} \right]^{\frac{1}{2}}}.
\end{eqnarray}
Here $\theta$ is the angle between the direction of the incident wave and the Zenith.

In Figure (\ref{fig:deviation angle}), we showcase the refraction effects of the F-layer. The {\it left} panel depicts the refractive angle $\delta \theta$ as a function of frequency, where each line corresponds to different zenith angles of the incoming light. The refraction angle is zero when observed directly from the zenith and gradually increases as one moves towards the horizon.
Similarly, in the {\it right} panel, we also demonstrate the refractive angle as a function of zenith angle for selected frequency channels.

The refractive effect could then be described as the change of coordinate of sky $T_{\rm sky}$
\begin{eqnarray}
\hat{T}_{\rm sky}(\nu,\theta,\phi)=T_{\rm sky}(\nu,\theta + \delta \theta,\phi),
\end{eqnarray}
where $\theta$ is altitude here and $\phi$ is azimuth. Due to the refraction effect, incoming radiation undergoes a tilting towards the zenith, result in additional antenna temperature rising which also varies with frequency, thus introduces a chromatic effect (Figure \ref{fig:deviation angle}), modifying the frequency dependency of both foreground radiation and red-shifted 21-cm signals. This chromatic alteration can introduce changes in the spectral characteristics of the observed radiation and signals, highly impacting data analysis.
The refractive effect caused by a single layer depends on the incident inclination angle of radio radiation and the variation with height of the index of refraction of the layer, which determined by the electron density and the frequency of the radio wave.

\subsection{Absorption}

Due to its lower altitude, the D-layer has a higher density of neutral gas molecules, resulting a higher collision frequency $\nu_c$ of electrons. Since the electron density $N_D$ in the D-layer is closely related to the neutral gas density, one could obtain the following  empirical formula \citep{shklovskii1960cosmic,evans1968radar}
\begin{eqnarray}
\nu_c = 3.65\frac{N_D}{T^{3/2}}\left[19.8+\ln \left(\frac{T^{3/2}}{\nu} \right) \right] ~ {\rm Hz}, 
\label{equation: electron collision frequency}
\end{eqnarray}
with $T$ representing the absolute temperature of the D-layer plasma, characterizing the thermal kinetic energy per particle. We have chosen a typical value of $200\ {\rm K}$, in agreement with the IGRF-13 
model\footnote{https://ccmc.gsfc.nasa.gov/models/IGRF$\sim$13/} of the International Reference Ionosphere project (IRI-2016)\footnote{http://irimodel.org}.
Here, $N_D$ is measured in electrons per cubic meter, and $\nu$ in {\rm Hz}. 

\begin{figure}
    \centering
    \subfigure{
\includegraphics[width=1\columnwidth]{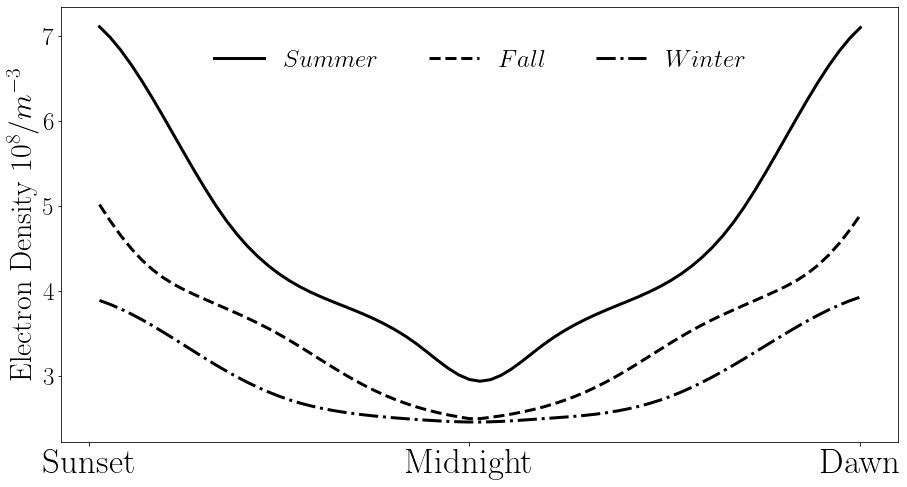}}
    \caption{Examples of nightly variation of electron number density in the D-layer during three specific nights across summer (Jun. 15), fall (Sep. 15) and winter (Dec. 15) at the observation site of $90^\circ 48' 24'' $E, $44^\circ 9'10''$N.
    We created the mock data under the assumption that observations are made in the fall, with a nightly minimum electron number density $N_D$ of $2.538 \times 10^8\  {\rm m}^{-3}$ and a maximum of $5.086 \times 10^8\ {\rm m}^{-3}$. These values were calculated using the IRI (2016) model.}
    \label{fig:Dynamic NeD}
\end{figure}

In this paper, we model the  D-layer electron density, $N_D$, as a function of time at our observation site. For our analysis, we selected a period of seven consecutive, relatively calm nights, illustrating the variability of $N_D$ across three days in three different seasons as examples in Figure (\ref{fig:Dynamic NeD}). As we can seen, $N_D$ is significantly affected by the altitude of the sun, displaying a decrease after sunset followed by an increase after midnight. In addition, $N_D$ shows clear seasonal variations as well, with values in summer approximately twice as high as those in winter. In the following, we will generate mock data, assuming observations are made during the fall season. 
With above specific values of $T$ and $N_D$, we find that the collision frequency $\nu_c$ varies between $2.95$ to $3.4\ \MHz$. This is lower than the value of $10\ \MHz$ quoted  by \cite{Shen21}, which was an approximate interpolation from rocket pressure measurements \citep{NICOLET1953200} at the median D-layer height of $75~ {\rm km}$. Our calculated collision frequency roughly corresponds to the value at approximately $80~ {\rm km}$ in the same measurement \citep{NICOLET1953200}.
Considering the substantial spatial and temporal variation in collision frequency and the uncertainties in such measurements, we do not consider this discrepancy as significant.

\begin{figure*} 
    \centering
    \subfigure{
    \includegraphics[width=17cm]{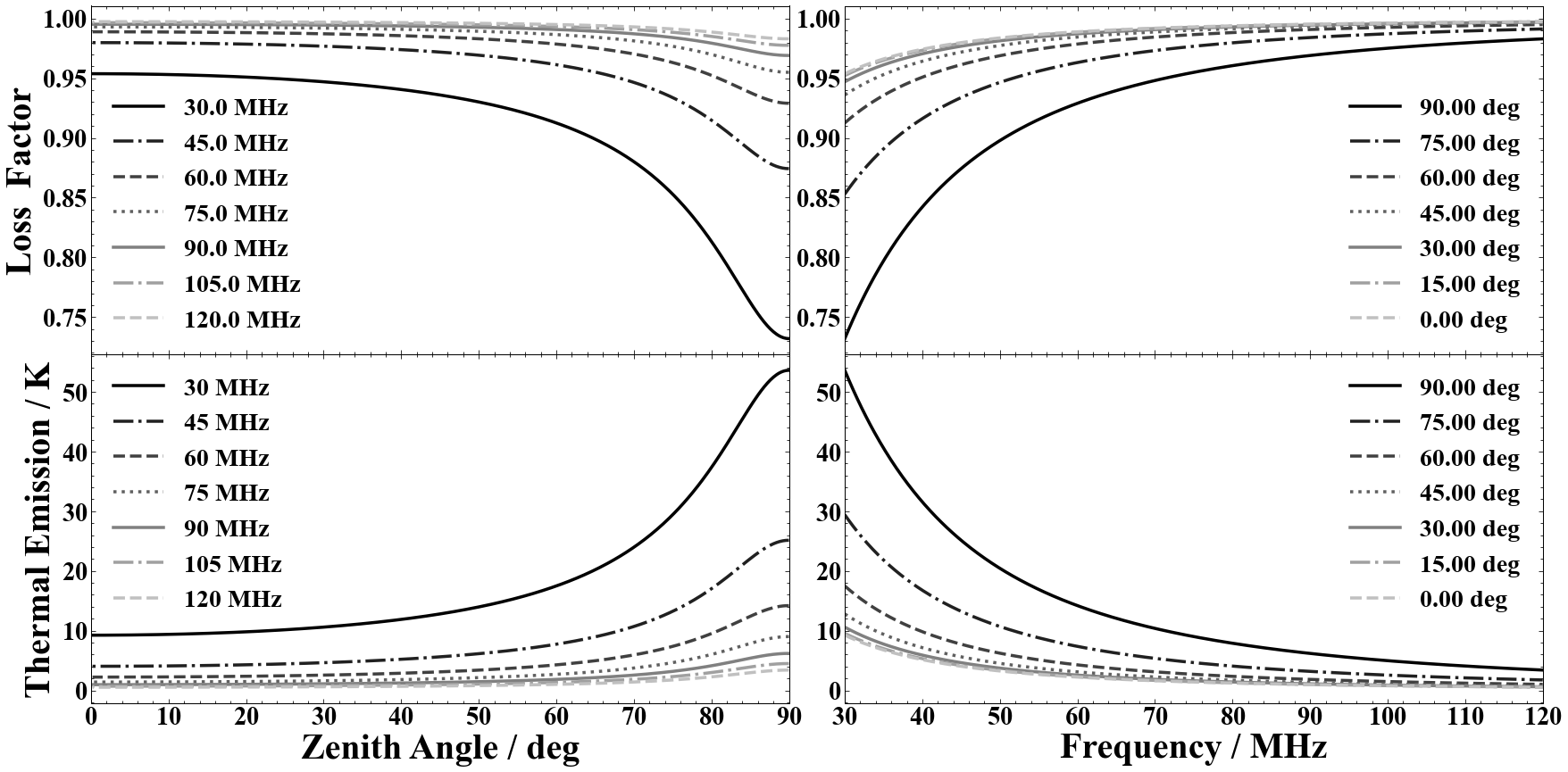}}
    \caption{ 
    Demonstration of absorption ({\it upper}) and emission ({\it lower}) of D-layer, with the electron density $N_D$ set to a typical value $2.5 \times 10^8 ~m^{-3}$.
    The {\it upper-left} panel depicts the absorption loss factor as a function of zenith angle across various frequencies. As illustrated, the absorption effect remains relatively constant for higher frequencies but decreases markedly at the horizon for lower frequencies.
    The {\it upper-right} panel displays the absorption loss factor as a function of frequency. As depicted, the absorption introduces a strong frequency dependence below $50\ \MHz$.
    In the {\it lower-left} panel, thermal emission is illustrated as a function of zenith angle. The emission gradually increases as the angle moves away from the zenith, reaching its maximum at the horizon. Finally, the {\it lower-right} panel shows thermal emission as a function of frequency. }
    \label{fig:loss factor}
\end{figure*}

Consequently, in D-layer, we can focus on the imaginary part of Equation (\ref{equation: simplified Appleton–Hartree equation}) and approximate the refractive index as
\begin{eqnarray}
    \eta_D \approx -\frac{1}{2}\frac{(\nu_c/\nu)\nu_p^2}{\nu^2+\nu_c^2}.
\label{equation: img part of sAH}
\end{eqnarray}
In a homogeneous ionospheric layer, the electric field can be described as a plane wave
\begin{eqnarray}
E(\Delta s)=E_0 \exp\ \left({-i\frac{2\pi \nu \eta}{c} \Delta s}\right),
\end{eqnarray}
where $c$ represents the speed of light in free space, $\Delta s$ is the path length along the direction of propagation within the ionosphere, and $E_0$ represents the initial electric field when $\Delta s$ equals to zero.
Since $\eta$ is imaginary, it produces an 
exponential attenuation of the wave, resulting in absorption \citep{Ved14}.
Given that the intensity of an electromagnetic wave is directly proportional to the square of its amplitude, we define the loss factor $\mathcal{L}$ as the remaining part after being absorbed by the ionosphere
\begin{eqnarray}
    \mathcal{L}(\nu,\theta)= \left | E\left ({\Delta s}(\theta) \right) \right |^2 =\exp\ \left({\frac{4\pi \nu \eta_D}{c} \Delta s(\theta)}\right),
\label{equation: Loss factor}
\end{eqnarray}
note $\eta_D<0$ in this definition.
The path length of an electromagnetic wave propagating in the ionosphere at a given incident angle $\theta$ have approximate expression derived from the model geometry \citep{Ved14}:
\begin{eqnarray}
    \Delta s(\theta) \approx \Delta H_D \left(1+ \frac{H_D}{R_e}\right)\left(\cos^2{\theta}+\frac{2H}{R_e}\right)^{-\frac{1}{2}}.
\end{eqnarray}
In this equation, $H_D$ represents the height of the D-layer, which ranges from $60\ \mathrm{km}$ to $90\ \mathrm{km}$, and $\Delta H_D$ is the width of the D-layer. The radius of the Earth, denoted as $R_e$, is approximately $6378\ \mathrm{km}$.

In the upper panels of Figure \ref{fig:loss factor}, we illustrate the loss factor, denoted as $\mathcal{L}$, in relation to the Zenith angle $\theta$ and frequency $\nu$. Specifically, the upper-left panel depicts $\mathcal{L}(\theta)$ curves across various frequencies. At lower frequencies, the loss remains relatively constant for rays approaching the Zenith, only to decrease gradually upon reaching a critical angle. For frequencies exceeding approximately $100\ \MHz$, absorption can be disregarded across all angles. Similarly, the upper-right panel showcases $\mathcal{L}(\nu)$ curves for specific Zenith angles. As indicated, wave absorption from the horizon can be roughly $5$ times greater than that of waves originating from the Zenith.
As shown, the loss factor stands at $0.73$ for a horizontally arriving ray at $30\ \MHz$, and $0.81$ at $40\ \MHz$. 
Due to the different ionospheric parameters utilized in this study, these outcomes show slight numerical difference from results in \cite{Ved14} and \cite{Shen21}, where the loss factors were approximately \rev{ $\sim 0.9$ and $\sim 0.6$} for a horizontally arriving ray at $40\ \MHz$, respectively. Considering the variability and uncertainties of these parameters, we do not consider such differences to be significant.

\subsection{Emission}
In addition to absorption, the D-layer also emits radio waves, thereby introducing additional thermal noise \rev{\citep{Datta2016}}. 
\citet{PAWSEY1951261} offers a method to compute the thermal emission, denoted as $T_{\rm te}$, in terms of equivalent temperature derived from the power absorption coefficient
$\chi=(4\pi \nu \eta/c) \Delta s(\theta)$:
\begin{align}
    T_{\rm te}(\nu,\theta) = \int_{H_{\rm Dmin}}^{H_{\rm Dmax}} \chi T_{\rm e}   \exp\left(-\chi \right)\ dH 
    = T_{\rm e}\left (1-\mathcal{L}(\nu,\theta) \right)
\label{equation: thermal emission}
\end{align}
Here $H_{\rm Dmax}$ and $H_{\rm Dmin}$ represent the upper and lower boundaries of the D-layer, respectively. $T_e$ denotes the electron temperature of the D-layer, with a typical value of $\sim 200\ \mathrm{K}$ as determined by the International Reference Ionosphere-IRI (2016) using IGRF-13 coefficients. 
In the lower panels of Figure (\ref{fig:loss factor}), we depict the emission temperature in relation to the Zenith angle and frequency. Analogous to absorption, the impact is more pronounced at lower frequencies and for horizontal waves. As shown, the peak value approaches approximately $50 \mathrm{K}$.

\begin{figure}
    \centering
    \subfigure{
\includegraphics[width=0.95\columnwidth]{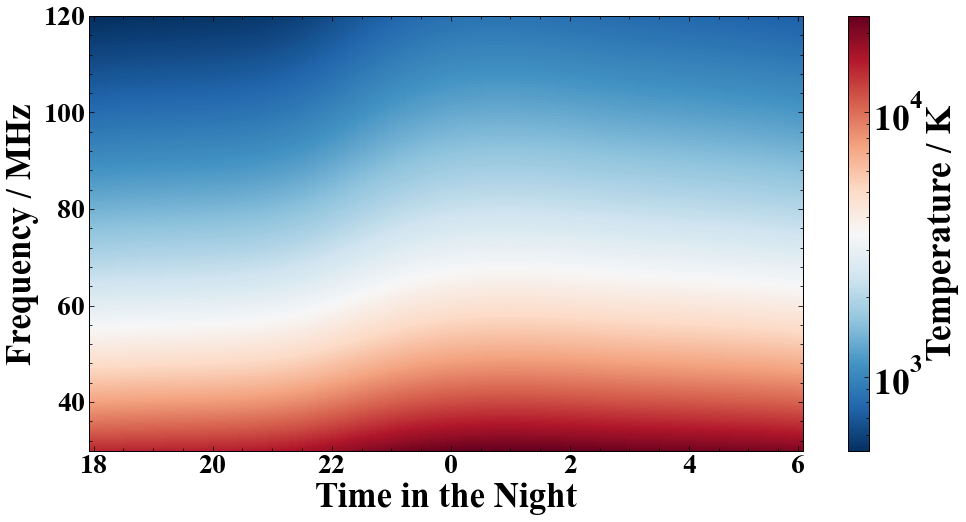}}
    \caption{The time-frequency waterfall plot of mock observation data, the values of the ionospheric parameters are taken from the \rev{IRI model as of September 2021.} 
    }
    \label{fig:time_order}
\end{figure}

Combining the ionospheric refraction, absorption and emission effects discussed earlier, we can  describe the antenna temperature, $T_A$, observed by the receiver as
\begin{eqnarray}
    T_{\rm A}(\nu) &=& \int_0^{2\pi}d\phi\int_0^{\pi/2} B(\nu,\theta,\phi)\bigl[ T_{\rm te} (\nu,\theta) +
    \mathcal{L}(\nu,\theta) \nonumber \\
    & & \times ~~ \widehat{T}_{\rm sky}(\nu,\theta,\phi)\bigr ]
    \sin{\theta}d\theta.
\label{equation: Antenna temperature}
\end{eqnarray}
Here $B(\nu, \theta, \phi)$ represents the beam function, as detailed in Section \ref{sec:inst_beam}. The angles $\theta$ and $\phi$ are defined in the local instrumental frame. 
It is noteworthy that Equation (\ref{equation: Antenna temperature}) differs from derivations like those in \citet{Ved14,Shen21}. In those works, refraction effects were described as a modified beam $B(\nu, \theta -\delta \theta, \phi)$. 
It is easy to notice that, by changing the variable $\theta +\delta \theta \to \theta^{\prime}$ and omitting the emission term $T_{\rm te}$, the expression in Eq. (\ref{equation: Antenna temperature}) bears a close resemblance to that of \citet{Ved14,Shen21} with the exception of the term $\sin \theta~ d\theta$ and the integral boundaries. Given that F-layer refraction occurs prior to absorption, emission, and reception at the antenna, our formula more closely aligns with the actual physical process.
As illustrated in Figure (\ref{fig:deviation angle}), the most significant deviation occurs near the horizon, where $\delta \theta \sim 2^{\circ}$.

\section{Foreground Fitting and Subtraction}

The success of the global spectrum experiment depends crucially on the effective removal of the overwhelming foreground contamination. This task is in some sense more challenging  than in the case of 21cm intensity mapping, for the observables are fewer, and the signal more smooth, and the range of available techniques for global spectrum measurement is more limited, one has to rely mainly on template fitting methods. In this paper, we evaluate three different fitting templates. The first two methods utilize distinct fitting formulas: the logarithmic polynomial fitting, and the physically motivated EDGES fitting formula. The last one involves templates that are customized to specific beam and ionosphere models. These templates are constructed using principal component analysis (PCA) of mock observational data.

\begin{figure*} 
    \centering
    \includegraphics[width=1\textwidth]{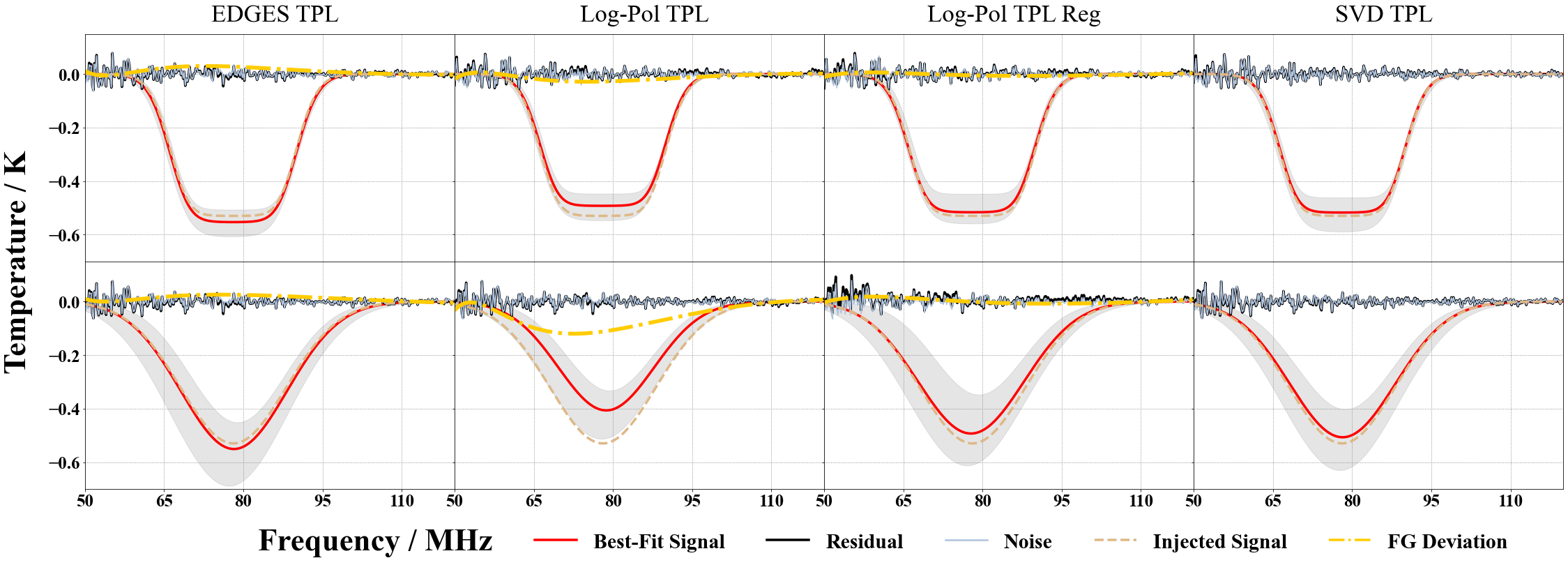}
    \caption{  \label{fig:fit_fiducial}
    Fitting results of mock observation data with ionospheric effect and fiducial beam model using different fitting templates. The first column shows the results using the EDGES foreground template (TPL), while the second and third columns display a fourth-order logarithmic polynomial model, both without and with higher-order regularization term, as described in Equation \ref{equation: likelihood_reg}.  The last column features the model-specific SVD templates. Two rows exhibit different variants of the injected 21cm signal. The upper panel demonstrates the flattened Gaussian model, while the lower panel represents the Gaussian model. 
    In each panel, the best-fitted 21cm signal is represented by a solid red line, while the actual injected signal is shown with an orange dashed line. The faint grey line demonstrates the residual between the data and the best-fit model, while the injected thermal noise is represented by the black line. \rev{The grey band surrounding the best-fitted signal indicates the $68.3\%$ credibility region, derived from the multi-dimensional distribution of MCMC samples.} 
    The yellow dash-dotted line illustrates the `foreground deviation', defined as the difference between the best-fitted foreground model in each panel and the identical model when fitted solely to the pure foreground. A deviation from zero, such as observed in the second panel of the second row, signifies overfitting by the template.
    As depicted, the flattened Gaussian signal model generally ensures a good fit across different foreground templates. Meanwhile, both the EDGES and the SVD template yield accurate fit for both signal models. 
    }
\end{figure*}

\subsection{Polynomial Fit Templates}
Given that for all major sources of the foregrounds, galactic synchrotron, free-free radiation and extra-galactic point sources \citep{Shaver99}, the spectrum is a smooth function of frequency and can typically be approximated by power laws with a running spectral index \citep{Petrovic2011}, the most straightforward foreground templates is some form of polynomial function. However, the specific form of this function can significantly impact the effectiveness of the foreground subtraction process.

Several studies have explored the application of polynomial functions for characterizing foregrounds. \citet{Oli08} demonstrated that a log-log function is capable of fitting the spectral dependency at individual pixels. \citet{Pritch10} assessed the efficacy of logarithmic polynomial fits up to the ninth order, concluding that at least the third order polynomial is necessary for adequate foreground removal. 
Subsequent studies \citep{Harker11,Liu12} have also confirmed that a third order logarithmic polynomial can recover the parameters well in this frequency range. 
However, the situation becomes more complex with additional frequency dependencies introduced by ionospheric effects. \citet{Shen21} demonstrated that a fifth-order polynomial is required in this situation. In our study, we consider a fourth order logarithmic polynomial to fit the foreground
\begin{eqnarray}
\log T_{\rm pol}(\nu) = \sum_{i=0}^{N=4} a_i \left[ \log \left(\frac{\nu}{\nu_c} \right) \right]^i + T_{21} (\nu)
\label{equation: log-fit}
\end{eqnarray}
Here, $\nu_c$ is the center frequency,  the cosmological signal $T_{21}$ is fitted using both flattened Gaussian (Eq. \ref{eqn:sig_flatgauss}) and Gaussian model (Eq. \ref{eqn:sig_gauss}).

\citet{EDGES2018Nature} proposed a physically motivated fitting formula, which is designed to capture various factors including the foreground, ionospheric effects, and specific instrumental influences, e.g., chromatic beams or minor calibration errors. The formula, centered at frequency $\nu_c$, is given by
\begin{eqnarray}
\label{equation:edges_foreground1}
T_{\rm EDGES}(\nu) &=& b_0 \exp\left[-b_3\left (\frac{\nu}{\nu_c} \right)^{-2}\right] \left( \frac{\nu}{\nu_c}\right)^{-2.5+b_1+b_2 \log(\nu/\nu_c)}\nonumber \\
&& + ~b_4\left(\frac{\nu}{\nu_c}\right)^{-2}
\end{eqnarray}
Here, the index value of $-2.5$ corresponds to the typical spectral index of the foreground. The parameter $b_0$ quantifies the overall amplitude of the foreground. Meanwhile, $b_1$ enables an adjustment to this typical spectral index, which varies across the sky. Additionally, $b_2$ captures any potential higher-order contributions. 
The ionospheric effects is encapsulated by parameters $b_3$ and $b_4$, which account for both absorption and emission effects. At the linear order, above formula could be approximated as
\begin{eqnarray}
    T_{\rm EDGES}(\nu) \approx a_0\left( \frac{\nu}{\nu_c}\right) ^{-2.5}+a_1\left( \frac{\nu}{\nu_c}\right)^{-2.5}\log\left( \frac{\nu}{\nu_c}\right) \nonumber \\
    +a_2 \left( \frac{\nu}{\nu_c}\right)^{-2.5} \left[\log \left( \frac{\nu}{\nu_c}\right)\right]^2+a_3\left( \frac{\nu}{\nu_c}\right)^{-4.5}+a_4\left( \frac{\nu}{\nu_c}\right)^{-2}
\label{equation: edges_foreground2}
\end{eqnarray}
To assess the effectiveness of different fitting methods, in our study, we employ both of the aforementioned formula to fit and subtract foreground contamination.

In this study, we employ the Monte Carlo Markov Chain (MCMC) method to fit the foreground and cosmological signal,  \rev{using the $\mathtt{emcee~ python}$ package \citep{Foreman_Mackey_2013}}. 
Several studies, such as those referenced in \citep{Shen21}, have employed the least-square fitting method, \rev{utilizing tools like the  $\mathtt{scipy.optimize.curve\_fit}$ routine from the  $\mathtt{SciPy}$ package \citep{Foreman_Mackey_2013}}. 
With some testing, we have found that the Monte Carlo Markov Chain (MCMC) method generally surpasses the least-squares fitting approach in performance. Notably, we observed that the least-squares fit can sometimes be especially sensitive to the initial guesses of parameters.
Given the posterior distribution of parameters provided by MCMC method, we have greater confident with the accuracy of the fit.

In the first two columns of Figure \ref{fig:fit_fiducial}, we showcase the fitting results. The \rev{second column displays} the logarithmic polynomial fitting, as described by Equation \ref{equation: log-fit}, and the first column features the EDGES template fitting, detailed in Equation \ref{equation: edges_foreground2}. Each row corresponds to a different variant of the injected 21cm signal models: the flattened Gaussian model is represented in the upper panels, and the Gaussian signal model is shown in the lower panels. 
Within each panel, the best-fitted 21cm signal is represented by a solid red line, while the  actual injected signal is shown with an orange dashed line. The grey line demonstrates the residual between the data and the best-fit model, while the injected thermal noise is represented by the black line. \rev{The grey band surrounding the best-fitted signal in each panel indicates the $68.3\%$ credibility region derived from the multi-dimensional distribution of MCMC samples, as shown in Figure \ref{fig:corner_plot1} and \ref{fig:corner_plot2}. }


\begin{figure*}
    \centering
    \subfigure{
    \includegraphics[width=1.8\columnwidth]{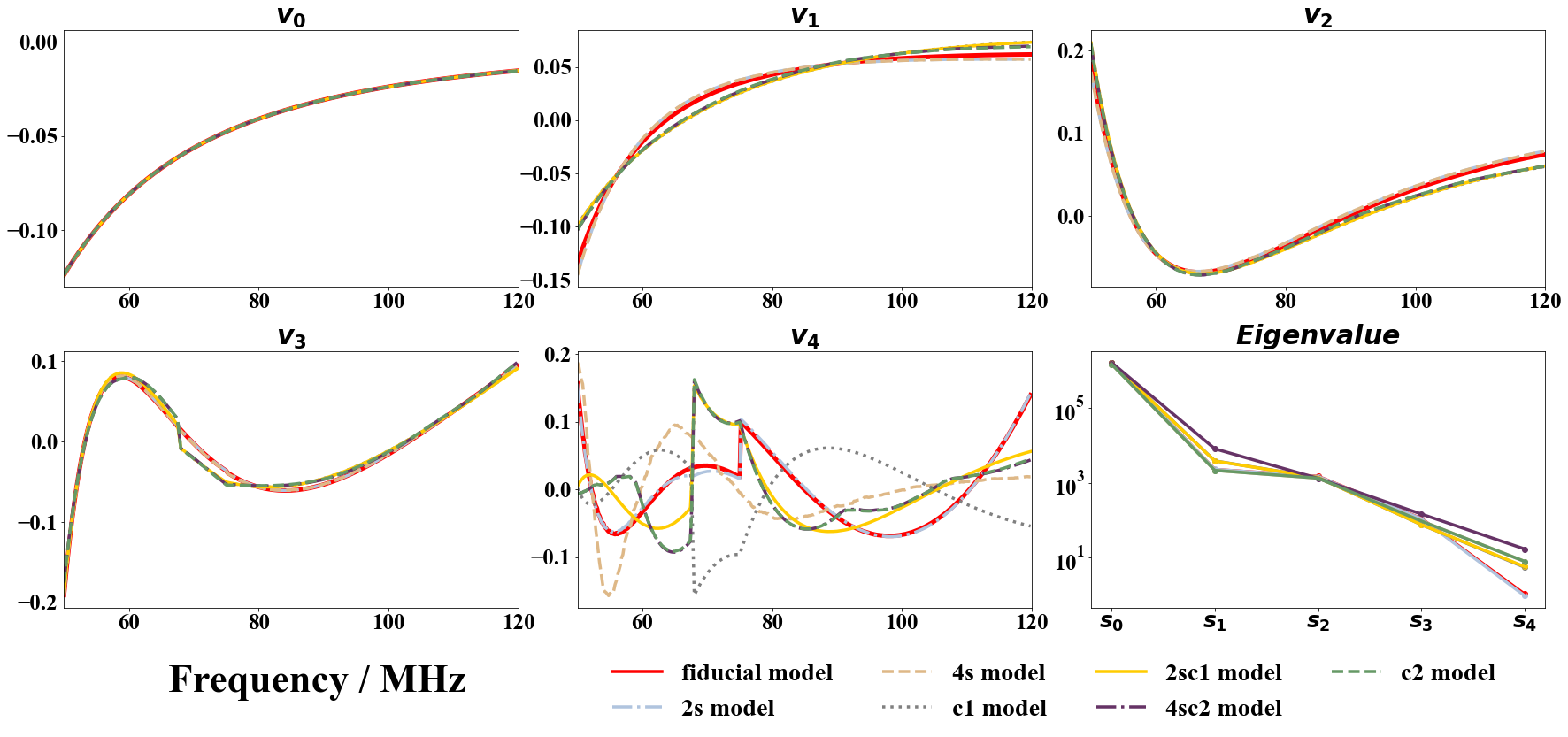}}
    \caption{The first five foreground SVD modes, ranging from $\mathbf{v}_0$ to $\mathbf{v}_4$. Within each panel, the thick solid red line illustrates the SVD mode corresponding to our fiducial beam model. Various other line styles represent the different beam distortion models, as detailed in Table \ref{table: beam_distortion}.
    The final panel displays the eigenvalues associated with these first five SVD modes.  \rev{Notice that various distorted eigenmodes presented here are solely for illustrative purposes. In the actual SVD template fitting, we continue to employ the fiducial SVD template to fit the distorted mock data.}
    }
    \label{fig:svdmodes}
\end{figure*}

Our analysis demonstrates that the more sophisticated EDGES fitting formula (the {\it first} column) generally outperforms the logarithmic polynomial models (the {\it second} column) in terms of fitting accuracy. Using the EDGES fitting template, the discrepancies between the best-fitted results and real signals are quite small for both Gaussian and flattened Gaussian signal models. In contrast, the logarithmic polynomial fitting does achieve fairly accurate fits for the flat-Gaussian signal, but its performance becomes less effective for the Gaussian signal model. 
This may be attributed to the relative simplicity of the Gaussian form, which makes it more susceptible to confusion and misfitting with polynomial template.
It is also noteworthy that even in the cases where the best-fitted signal substantially differs from the actual signal, e.g. in \rev{the lower panel of the second column}, the residual ({\it black line}) remain closely aligns with the thermal noise. 
This suggests the presence of significant internal degeneracy between the foreground model and the signals, leading to overfitting. This is further demonstrated by the `foreground deviation', represented as the yellow dash-dotted line. This deviation is defined as the difference between the best-fitted polynomial foreground model, which is specifically the foreground in Equation (\ref{equation: log-fit}) with the fitted parameters $a_i$, and the same polynomial model when fitted exclusively to the pure foreground. As observed, this deviation aligns with the signal shift illustrated in the same panel.

\rev{To further understand the details of our fitted result, we display the multi-dimensional distribution of all model parameters for two representative examples. Specifically, Figure \ref{fig:corner_plot1} showcases the distribution using the EDGES foreground template paired with a flattened Gaussian signal model, while Figure \ref{fig:corner_plot2} presents the results from fitting a logarithmic polynomial with a Gaussian signal. 
For the former model, all parameters show a certain level of degeneracy. The distribution is approximately Gaussian, with the exception of the flattening factor $\tau$. This leads to an effective fit to the signal depicted in the first panel of Figure \ref{fig:fit_fiducial}, where the best fit model closely match the true signal with a relative narrow error band.
On the other hand, the distribution combining logarithmic foreground and Gaussian signal model (Figure \ref{fig:corner_plot2}) is quite different. As shown, the foreground coefficients $a_n$ of odd polynomial orders tend to be independent from those of even orders. However, within the subsets of even and odd coefficients, degeneracies persist.
Moreover, the signal parameters exhibit larger degeneracy and longer tails, particularly for amplitude $A_0$ and trough location $u$. Consequently, as illustrated in the second panel of the lower row, there is a noticeable shift in the best-fit curve and a broadened error band.
}

In \citet{Shen21}, the author demonstrated that even with a simpler foreground model, generated by extrapolating the Haslam $408\ \MHz$ map with a constant spectral index, a logarithmic polynomial fit to the Gaussian signal model could lead to considerably large residuals in some ionospheric models. Specifically, they showed that a fifth-order logarithmic polynomial could yield a biased best-fit model with a residual comparable to the cosmological signal when considering an ionosphere model with a constant D-layer. 
\rev{While our findings might appear to differ from those of \citet{Shen21} at first glance, given that our residuals in this model are significantly smaller, it is important to notice that our best-fit signal is also biased here. 
The reduced residuals in our case is likely due to a more comprehensive sampling process using the MCMC approach. This highlights the nuanced complexity of the fitting process.
}

To mitigate the overfitting for logarithmic polynomial model, we further add an extra regularization term to the logarithmic likelihood function, penalizing the higher-order polynomial terms, specifically
\begin{eqnarray}
    \ln \mathcal{L}_{\rm reg} = 
    -\sum_i^N \left ( \mathcal{R}_i {a_i}^2 \right)
\label{equation: likelihood_reg}
\end{eqnarray}
Here, we simply set the regularization parameters $\mathcal{R}=(0,0,10,10^2,10^3)$. 
As shown from the third column of Figure \ref{fig:fit_fiducial}, this regularization significantly improves the fitting results and effectively prevents overfitting.
\rev{We notice that the final conclusion are not significantly influenced by the choice of regularization parameters. The primary guideline for selecting above values is to ensure that higher order coefficients are greater than the lower order ones.
}


\subsection{Model-based SVD Template Fit}

In addition to polynomial fitting, alternative templates can also be employed for analysis. Other than given functional forms, one can construct templates with \rev{principal} components of (mock) observational data.
\rev{Principal} Component Analysis (PCA) is widely used in data analysis of many observations, notably in foreground removal of 21cm intensity mapping. Instead of blindly subtracting major modes as in 21cm intensity mapping, our focus here is on developing foreground templates that incorporate potentially complex signatures found in real observations. These include effects such as ionosphere, beam chromaticity, and spatially dependent foreground. 
With an accurate instrument and sky model, it is possible to decompose mock observations into distinct frequency modes to achieve this purpose. This methodology is in line with similar approaches previously discussed by \cite{Tauscher18}, where its application in global spectrum observation has been demonstrated.

To construct these foreground modes, we begin with a mock foreground-only observation datacube, denoted as $\mathbf{D}_{\rm fg}(t,\nu)$, which has dimensions of $n_{\rm t} \times n_{\nu}$, where $n_{\rm t}$ and $n_{\nu}$ represent the number of data points along the time and frequency axes, respectively.
Here we have incorporated all ionospheric effect shown previously in Equation (\ref{equation: Antenna temperature}). Following this, we then perform the singular value decomposition (SVD) of $\mathbf{D}_{\rm fg}(t,\nu)$
\begin{eqnarray}
\mathbf{D}_{\rm fg}=\mathbf{U}\mathbf{\Sigma} 
\mathbf{V}^T
\end{eqnarray}
Here, $\mathbf{\Sigma}$ is a diagonal matrix of singular values in descending order, matrix $\mathbf{U}$ has the dimension of $n_{\rm t} \times n_{\rm t}$ and $\mathbf{V}$ is of $n_{\nu}\times n_{\nu}$. $\mathbf{V}$ consists of right singular vectors $\mathbf{v}_i$, with the first few of which capturing the most significant  frequency characteristics of the `observed'
foreground, combining instrumental, and ionospheric effects. 
Using these foreground templates $\mathbf{v}_i$, we can then perform a fitting with the following equation:
\begin{eqnarray}
\label{eqn:svdfit}
    T_{\rm SVD}(\nu)=\sum_{i=0}^{N_{\rm SVD}} a_i \mathbf{v}_i + T_{21} (\nu) .
\end{eqnarray}

\begin{figure*}
    \centering
    \subfigure{
    \includegraphics[width=1\textwidth]{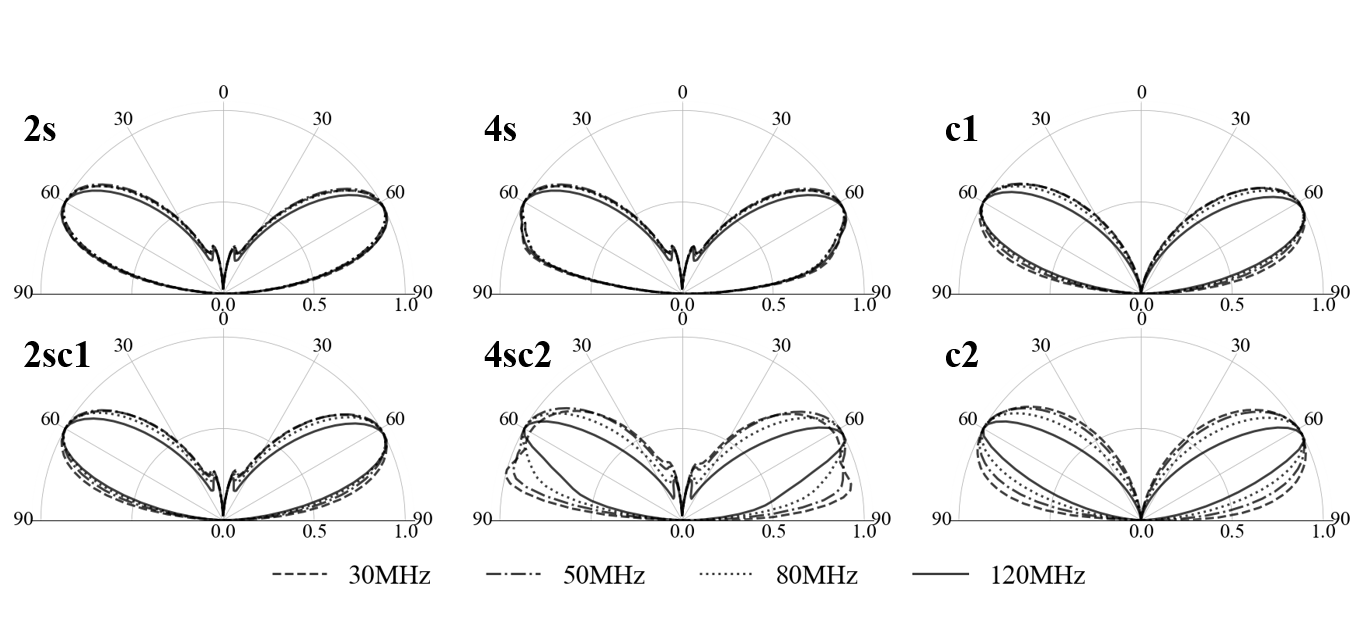}}
    \caption{    \label{fig:beam_full}
    Various beam distortion model considered in this paper, each with distinct characteristics. (1) $2s$ Model: This model adds two extra side lobes leaning towards the zenith, while the main lobes remain unchanged. 
    (2) $4s$ Model: Characterized by four additional side lobes, two of which merge with the main lobes, causing significant distortion of the main lobe pattern. 
    (3) $c1$ Model: Exhibits a relatively minor extra frequency dependency. 
    (4) $2sc1$ Model: Incorporates two additional side lobes together a relatively small frequency dependence.
    (5) $4sc2$ Model: Features four additional side lobes and substantial distortion of the two main lobes, coupled with relatively large frequency dependency. 
    (6) $c2$ Model: Similar to $c1$ model, but with a larger frequency dependency. 
    The exact functional forms of these distortion models are detailed in Table \ref{table: beam_distortion}. }
\end{figure*}

In Figure (\ref{fig:svdmodes}), we present the first five foreground modes. In each panel, the thick solid line denotes the SVD templates assuming fiducial beam model. It is evident that the first couple of modes are very smooth and featureless, with more complex structures starting to appear from higher order modes. 
Furthermore, the last panel illustrates the first five eigenvalues. Spanning roughly five orders of magnitude, these eigenvalues provide valuable insight into the relative importance of each mode within the fitting process.
In the last column of Figure (\ref{fig:fit_fiducial}), we display the fitting results using the first three SVD templates, i.e. $N_{\rm SVD}=2$. 
Compared to the polynomial fittings shown in the first two columns, exhibit significantly improved accuracy for both the flattened Gaussian and the Gaussian signal models, with minimal deviation from the injected signal.

\section{Beam Distortion}
\label{sec:beamdistortion}

In this section, we will focus on a crucial source of potential systematic errors: the beam distortion. Beam distortions that vary with frequency introduce additional spectral structures, potentially affecting the efficacy of the fitting process. Therefore, the main  goal here is to assess the robustness of various fitting templates against such distortions. In real-world scenarios, beam distortions can arise from various factors, including inaccuracies in beam modeling, manufacturing errors, physical deformities of the equipment, or frequency distortions at the digital processing stage.
Since the SVD templates are constructed based on specific models of the instrument, sky, and ionosphere, any deviations from these models could significantly alter the templates, thereby impacting the effectiveness of the fitting. \rev{To address this, we construct SVD templates in this section by assuming mock data with a fiducial beam model and ionospheric effects. We then use these fiducial SVD templates to fit mock data observed with various distorted beam models. This approach allows us to thoroughly assess the robustness of the SVD templates against such distortion.}

\begin{figure*}
\centering
\includegraphics[width=\textwidth]{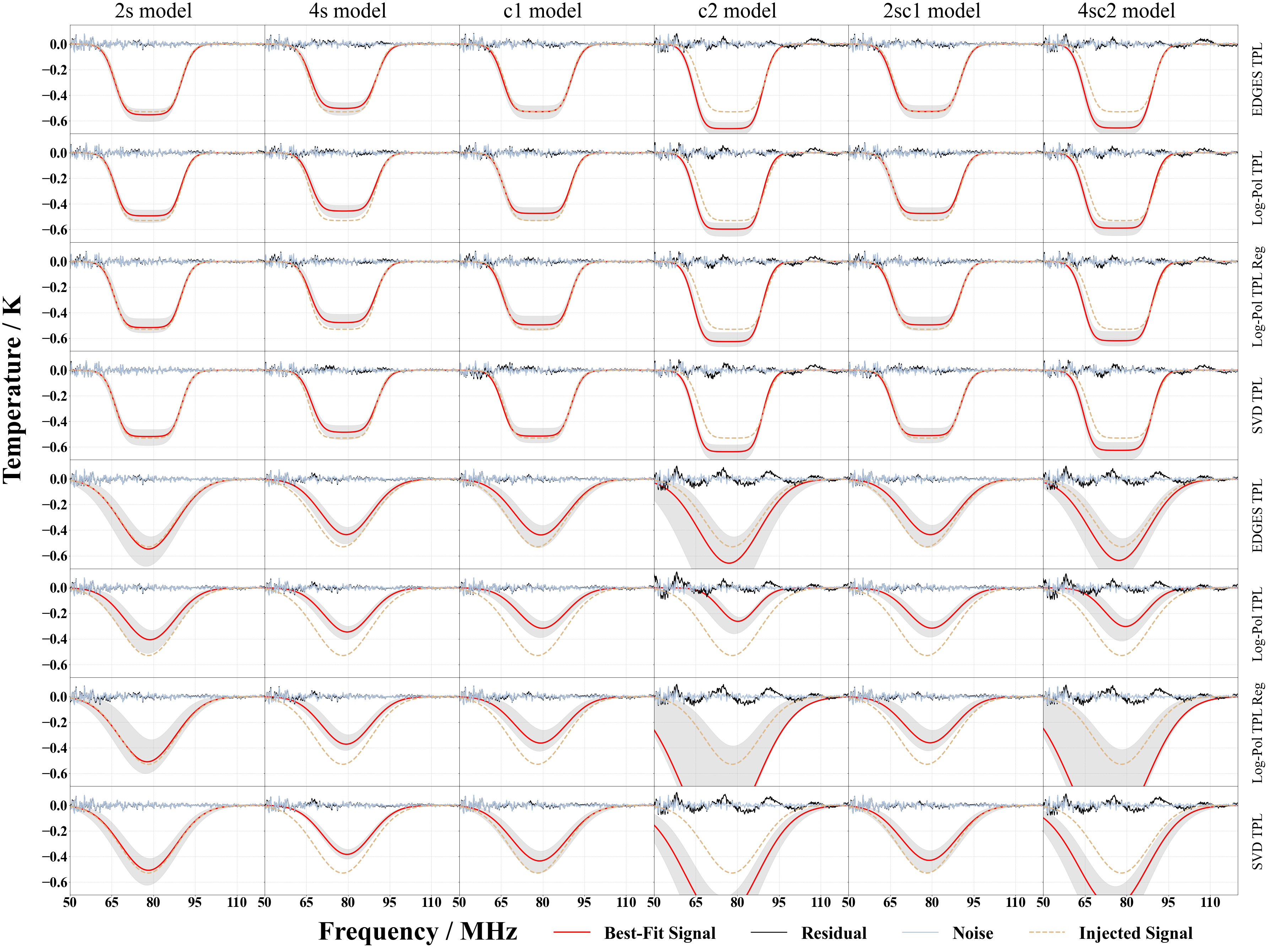} 
    \caption{  \label{fig:fit_beam_diff_all} 
    The fitting results of mock data observed using antennas with various distorted beams. Each column corresponds to a specific beam model, as illustrated in Figure \ref{fig:beam_full} and Table \ref{table: beam_distortion}. 
    The first four rows display results in which both the injected cosmological signal and the signal fitting template are based on the flattened Gaussian model, while the subsequent four rows utilize the Gaussian signal model. 
    The first and fifth rows present the best-fit results using the EDGES template. Similarly, the second and sixth rows illustrate the outcomes for logarithmic polynomial fitting without the regularization term, while the third and seventh rows incorporate  the regularization term (Equation \ref{equation: likelihood_reg}). 
    Finally, the fourth and eighth rows display the results obtained by fitting the first five SVD templates. The line styles and color schemes used here are the same with those in Figure \ref{fig:fit_fiducial}.  
    Notice that the SVD templates used here were constructed based on the fiducial beam model and subsequently fitted to mock data incorporating distorted beams.
    As shown, the flattened Gaussian model generally yields reasonably accurate fits, exhibiting only mild dependency on the beam distortion model and foreground template used. Among all the beam models considered here, the $4sc2$ model, which represents the most pessimistic scenario, performs the worst. The performance of the other distorted beam models varies depending on the chosen foreground and signal templates.
    }
\end{figure*}

In the subsequent analysis, we investigate two types of beam distortion: additional beam chromaticity and unaccounted sidelobes. For such purpose, we create several models that incorporate these distortions. 
Defining $\vartheta$ as the angle from the center of the main lobes (of which there are two in our case), we introduce a specific form of perturbation to our fiducial beam function $B(\vartheta,\nu)$, described as 
\begin{eqnarray}
B^{\prime}(\vartheta,\nu)=
B(\vartheta,\nu)\left[1+C(\nu)\sin \left( H(\nu) \frac{\vartheta}{2 \pi} \right) \right]
\label{equation:fwhm_change}
\end{eqnarray}
Here, $H(\nu)$ represents the Half Power Beam Width of the main lobe of our fiducial beam, and $C(\nu)$ is a frequency-dependent distortion coefficient. We investigate two variations of $C(\nu)$ in this mock experiment. 
The first model, $C_1(\nu)$, exhibits a linear variation from  $0.03$ at $30 ~\MHz$ to $-0.03$ at $120 ~\MHz$. The second model, $C_2(\nu)$, has a much larger effect, varying from $0.1$ at $30~ \MHz$ to $-0.1$ at $120~\MHz$.
For additional side lobes, we also consider two models. The first model, $SL_1 (\vartheta, \nu)$, features two sidelobes, each with an amplitude one-fifth that of the main lobes. The center of each sidelobe in this model is positioned approximately $11 \deg$ away from the pointing center. The second model, $SL_2(\vartheta, \nu)$, similarly includes two sidelobes, but these are located near the horizon, providing a distinct variation in beam structure for our analysis.

\begin{table}
    \centering {\large$B^{\prime}(\vartheta,\nu)$}
    \begin{tabular}{c|c|c}
         \hline\hline
         1& 2s Model & $B(\vartheta,\nu)+SL_1 (\vartheta,\nu)$\\ 
         \hline
         2& 4s Model & $B(\vartheta,\nu)+SL_1(\vartheta,\nu)+SL_2 (\vartheta,\nu) $\\
         \hline
         3& c1 Model & $B(\vartheta,\nu)\left[1+C_1(\nu)\sin\left( H(\nu) \frac{\vartheta }{2\pi}\right)\right]$\\
         \hline 
         4& c2 Model & $B(\vartheta,\nu)\left[1+C_2(\nu) \sin\left( H(\nu) \frac{\vartheta }{2\pi}\right) \right] $\\ [1ex] 
         \hline
         5& 2sc1 Model & $\left(B(\vartheta,\nu)+SL_1\right)\left [ 1+C_1(\nu)\sin\left( H(\nu) \frac{\vartheta }{2\pi}\right) \right ] $\\
         \hline
         6& 4sc2 Model & $\left(B(\vartheta,\nu)+SL_1+SL_2\right)\left [ 1+C_2(\nu) \sin\left( H(\nu) \frac{\vartheta }{2\pi}\right) \right]$ \\ 
         \hline\hline
    \end{tabular}
    \caption{The exact functional forms of each beam distortion model illustrated in Figure \ref{fig:beam_full}. Here $B(\vartheta,\nu)$ represents our fiducial beam model, the term $SL_1(\vartheta, \nu)$ and $SL_2(\vartheta, \nu)$ refer to two distinct side-lobe models, each centered at two different angles. Meanwhile, the parameter $C(\nu)$ controls the additional chromaticity introduced in the beam distortion.}
    \label{table: beam_distortion}
\end{table}

Incorporating both the extra frequency-dependency and side lobes, we consider a total of six distinct models characterizing the beam distortion from our fiducial model. The names and specific functional forms of these models are detailed in Table \ref{table: beam_distortion}, with their corresponding shapes illustrated in Figure (\ref{fig:beam_full}). 
From the figure, the $2s$ and $4s$ models introduce a sidelobe and modify the beam's shape, yet they do not significantly change the frequency dependency. In contrast, the $c1$ and $c2$ models adjust the width of the main lobe as a function of frequency to emulate an enhanced chromaticity, without introducing any additional angular structures. Finally, we also introduce two more complex models $2sc1$ and $4sc2$, which combine both effects. Consequently, these models represent more severe beam distortions, with the $4sc2$ model depicting the most extreme distortion scenario considered in this paper.

\begin{figure}
\centering
\includegraphics[width=\columnwidth]{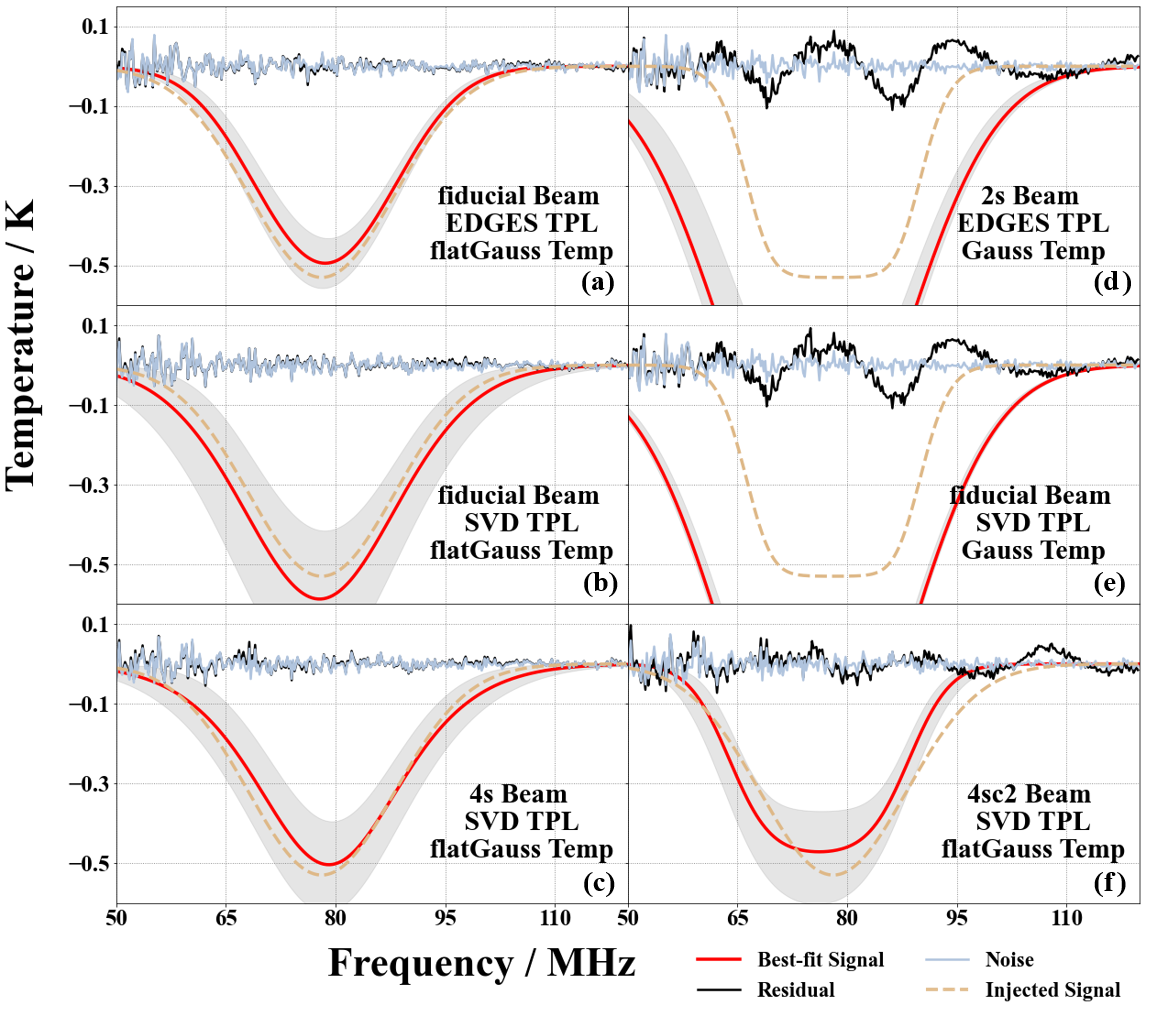} 
    \caption{    \label{fig:crossfit}  
    The demonstration of flattened Gaussian template fit to Gaussian signal model (the first column and the last panel), and vice versa (the first two panels in the second column). 
    In the first column, the top two panels (panel a and b) display the results with fiducial beam model and fitted using the EDGES and SVD templates, respectively. The bottom two panels (c and f) demonstrate results for the $4s$ and $4sc2$ beam models using the fiducial SVD template. 
    In all four instances, the fit to the injected signal is reasonably accurate, due to a small flattening factor $\tau$ being a good approximation to the Gaussian model.
    Conversely, the Gaussian signal template fitting to the flattened Gaussian signal is presented in the first two panels (d and e) of the second column. Here, even with the fiducial beam model, the best-fitted signal shows a significant deviation from the injected signal. 
    }
\end{figure}

The SVD templates derived from mock data using various distorted beam models have been shown in Figure \ref{fig:svdmodes}, 
Notably, the first mode $\mathbf{v}_0$, primarily influenced by the foreground itself, remains stable for all beam models. Minor deviations are observed in modes $\mathbf{v}_1$ to $\mathbf{v}_3$. The most significant change, however, is evident in the fifth mode $\mathbf{v}_4$, where the impact of beam distortion is most noticeable.  
Given that the fifth eigenvalue is over five orders of magnitude smaller than the first, this might appear contradictory to the drastic beam shape change introduced in these models. It is important to recall, however, that these SVD templates are decompositions of the mock global spectrum datacube, the major components of which are the foreground itself. 
Moreover, in the last panel of the figure, there is a noticeable shift in the significance of each component's eigenvalues. Most notably, the eigenvalues of the second and fifth modes display the greatest variation, differing by a factor of a few. This variation highlights the impact of beam distortions on the relative importance of different SVD components in our analysis. 
\rev{Notice that in the actual SVD template fitting, we continue utilizing the fiducial SVD template to fit the mock data affected by the distorted beams. }

In Figure \ref{fig:beam_full}, we show the results of all beam models, fitted by various foreground templates and signal models. 
Each column corresponds to the mock data observed by individual distorted beams. Within each row, this mock data is then fitted with different templates and signal models. Specifically, the first four rows focus on the flattened Gaussian signal model, while the last four rows are dedicated to the Gaussian signal model. 
From the figure, we first notice that the  choice of the signal model plays a crucial role in the performance. Notably, the flattened Gaussian model, illustrated in the first four rows, achieves a good fit in almost all scenarios with the exception of the most challenging $c2$ and $4sc2$ beam model. This indicates its robust performance across most situations. In contrast, the performance with a Gaussian signal, as shown in the last four rows, varies depending on the specific beam distortion model and the selected foreground template.

To determine whether such differences in performance originate from the modelling or the fitting template of the signal, further analysis is required. In Figure \ref{fig:crossfit}, we conduct a cross-fitting exercise, which is performed using a flattened Gaussian/Gaussian template to fit a injected Gaussian/flattened Gaussian profile, thereby allowing us to separate and examine the impact of the  modelling and fitting template of the signal. 
In the first two panels of the left column (panels a and b), we present the fitting results of a Gaussian signal using a flattened Gaussian template, combined with EDGES and SVD foreground models, and assuming a fiducial beam. As shown, the results here appear reasonably accurate in both situations. Even with distorted beam models, such as the $4s$ model shown in panel (c) and the $4sc2$ model in panel (f), the SVD foreground template employing a flattened Gaussian signal successfully extracts the injected Gaussian signal with minimal distortion. This is because the flattened Gaussian function approximates a Gaussian as the flattening factor $\tau$ approaches  zero. 
Conversely, however, the Gaussian function faces more challenges in extracting a flattened Gaussian signal, which is evident in panels (d) and (e), where both EDGES and SVD foreground fittings are applied. Therefore, above results show that the flattened Gaussian template offers more robustness across a variety of situations.

Besides the functional form of the signal model, we also examine the performance of various foreground templates against different distorted beam models. As shown in Figure \ref{fig:beam_full}, the physically motivated EDGES foreground template, displayed in the first and fifth rows, generally provides a better fit than the logarithmic polynomials, shown in the second and sixth rows. This is especially the case for the Gaussian signal, where the EDGES template generally outperforms the polynomial across all beam models.  However, for beam models that exhibit  greater chromaticity, specifically the $c2$ and $4sc2$ models, the logarithmic polynomial seems to provide a better fit for the flattened Gaussian signal. 
\rev{Furthermore, integrating the regularization term into the logarithmic polynomial fit, demonstrated in the seventh row, does improve the fitting results for mild beam distortion models (e.g., $2s$ and $4s$ models). Nonetheless, this approach does not yield improvements for models with more significant distortion.
}

Moreover, the SVD templates constructed from fiducial beam model fit the signal surprisingly well. This is particularly true for the less distorted beam models, that is, models other than the $c2$ and $4sc2$ models.
Notice that, here, we have utilized all five SVD modes in Figure \ref{fig:svdmodes} to fit the data. In most of these less distorted beam models, the SVD template ranks as the second-best fit, performing only slightly below the EDGES foreground model. 
However, for beam models with greater chromaticity, i.e. Model $c2$ and $4sc2$, the performance declines significantly, especially with Gaussian signal model. This is understandable, as the extra frequency dependence will substantially alters the decomposition, leading to larger errors. 
Consequently, as long as the real-world beam function does not introduce a substantial additional frequency dependence, the SVD template from theoretical beam model should still offer a robust fit to observational data.

From the figure, we also conclude that, among various beam distortions, the chromaticity is clearly the most crucial factor. This is evident in the poor performance of the $c2$ and $4sc2$ beam model. On the other hand, significant but achromatic changes to the beam shape do not adversely impact the fitting process, with both EDGES and SVD templates performing reasonably well. 
It is important to note, however, that this conclusion is based on the relative simple model incorporated in our mock observation. More detailed examination will be necessary in future studies to fully understand the process.


\section{CONCLUSION}
\label{conclusion}

In this paper, we study the challenges faced by the 21cm global spectrum experiment, with a specific focus on the impacts of the ionosphere and beam distortion. To address these challenges, we have developed a mock simulation pipeline that integrates ionospheric effects and a beam design modeled on an instrument currently operational for observations in Hami, Xinjiang, China.
With the extremely low signal-to-contamination ratio of the experiment,  the additional spectral structures introduced by Earth's ionosphere and chromatic beam variations can lead to systematic errors in the interpretation of the cosmological signal. The challenge is particularly acute in 21cm global spectrum observation, as most single-antenna experiments lack spatial resolution, and the cosmological signal is as smooth as the foreground. Therefore, it is crucial to assess the efficacy and robustness of various spectral fitting templates in this context.

In our approach to modeling ionospheric effects, we follow methodologies used in similar studies \citep{Ved14,Shen21}, including the refraction, absorption, and emission of incoming electromagnetic waves. We have made subtle improvements in calculating the refractive effect. 
Our ionosphere model incorporates a height-dependent but temporally-constant F layer, along with a height-independent but temporally-varying D layer. Key physical parameters within this model are calculated based on the local electron density, specific to the location of our instrument. These calculations are derived from the IGRF-13 model from the IRI-2016 project.
Consequently, there are minor differences in the refraction angle and absorption effects compared to studies like \cite{Shen21}. Considering the simplified ionosphere modeling in both our works, there is a clear need for more complex and realistic models in future studies.
Furthermore, consistent with previous studies, our analysis demonstrates that ionospheric effects are notably less significant near the zenith, suggesting a potential optimization for future instrument design that features a zenith-oriented main lobe.

Utilizing mock observational data incorporating aforementioned ionosphere and beam models, we proceeded to test various fitting templates. Specifically for the foreground, we conducted comparison among a logarithmic polynomial, the EDGES function, and a model-based SVD template. Regarding the cosmological signal, we compared a Gaussian with a flattened Gaussian model. 
Across all combinations of signal and foreground templates, we discovered that the EDGES function and SVD based template yield the best results. Meanwhile, applying a logarithmic polynomial fit to a Gaussian signal could result in overfitting the data, leading to errors in signal extraction.

The challenge of the SVD template, of course, lies in its requirement for precise modeling of all components of the mock simulation. 
To further assess the robustness of the various templates against systematic errors, we examined the impact of beam distortion. For this, we applied six models of beam distortion to the baseline beam and repeated the fitting process using all foreground and signal templates. Notably, the SVD templates were constructed specifically using the baseline beam model.
Our custom-designed beam distortion models aim to capture two distinct aspects of beam variation: additional chromaticity and achromatic angular structures.

The fitting results clearly indicate that chromaticity is the most significant factor impacting signal extraction. 
Beams with higher frequency dependence introduce additional structures that can lead to misfitting of the signal. In terms of templates, the combination of the EDGES foreground template and flattened Gaussian model proves to be quite robust in almost all scenarios, making it a preferred choice. Conversely, the logarithmic polynomial and Gaussian model combination is the least effective. Our cross-fitting exercise showed that while the flattened Gaussian template can reasonably extract an injected Gaussian signal, the reverse is not as effective.
On the other hand, the SVD template demonstrates surprising robustness. For beams with only mild distortion, the SVD template is capable of fitting the data reasonably well. However, as the chromaticity becomes more pronounced, the performance of the SVD template significantly deteriorates, making it less effective compared to other methods.

Finally, it is important to bear in mind that our ionosphere model remains somewhat oversimplified. For instance, the electron number density in the D layer, which influences the loss coefficient, also exhibits vertical structures, and the parabolic model used for electron number density in the F layer is clearly a very rough approximation.
Furthermore, other potential sources of chromaticity could affect the observations as well. For example, the structure of the horizon may have a significant impact \citep{pattison2023modelling}. Unpredictable space weather events, inaccuracies in receiver response, and other factors can introduce further complications.
Despite the challenges in achieving highly accurate models of the ionosphere and other systematic effects, evaluating potential systematic errors in a simplified setting is still valuable. This work serves as a foundation for such assessments, and we plan to pursue more detailed investigations in future studies.

\section*{Acknowledgements}
This research is supported by the National SKA Program of China (Grants Nos. 2022SKA0110200 and 2022SKA0110202), National R\&D program 2022YFF0504300, the NSFC grant 12361141814, and the CAS grant ZDKYYQ20200008.
Simulation results of ionosphere parameters have been provided by the Community Coordinated Modeling Center (CCMC) at Goddard Space Flight Center through their publicly available simulation services (https://ccmc.gsfc.nasa.gov).  The IRI 2016 Model was developed by the IRI Working Group.


\appendix
\section{Parameter Constraints}
    
    \begin{figure*}[h]
    \centering
        \subfigure{
        \includegraphics[width=\textwidth]{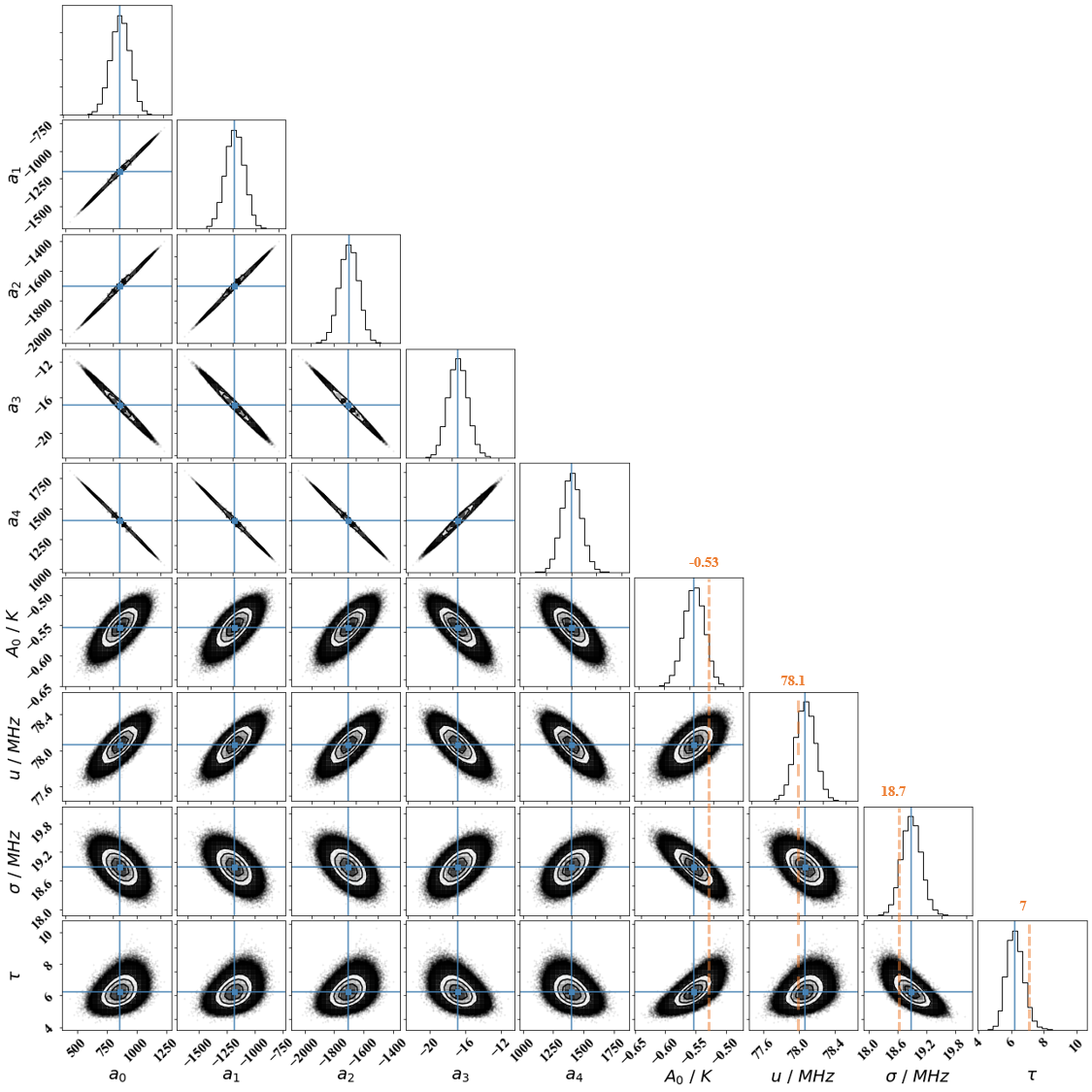}}
    \caption{ \label{fig:corner_plot1}
    \rev{The distribution of model parameters from the MCMC analysis, showcasing the results of the EDGES foreground template with a flattened Gaussian model. We used the Corner package \citep{corner} generating this plot.
    As illustrated, all parameters exhibit some degree of degeneracy, and the overall distribution is approximately Gaussian, with the exception of the flattening factor $\tau$ which displays a slightly longer tail. For the foreground parameters, specifically $A_0$, $u$, $\sigma$ and $\tau$, their true values are indicated by an orange dashed line. }}
    \end{figure*}
    
    \begin{figure*}[h]
        \subfigure{
        \includegraphics[width=\textwidth]{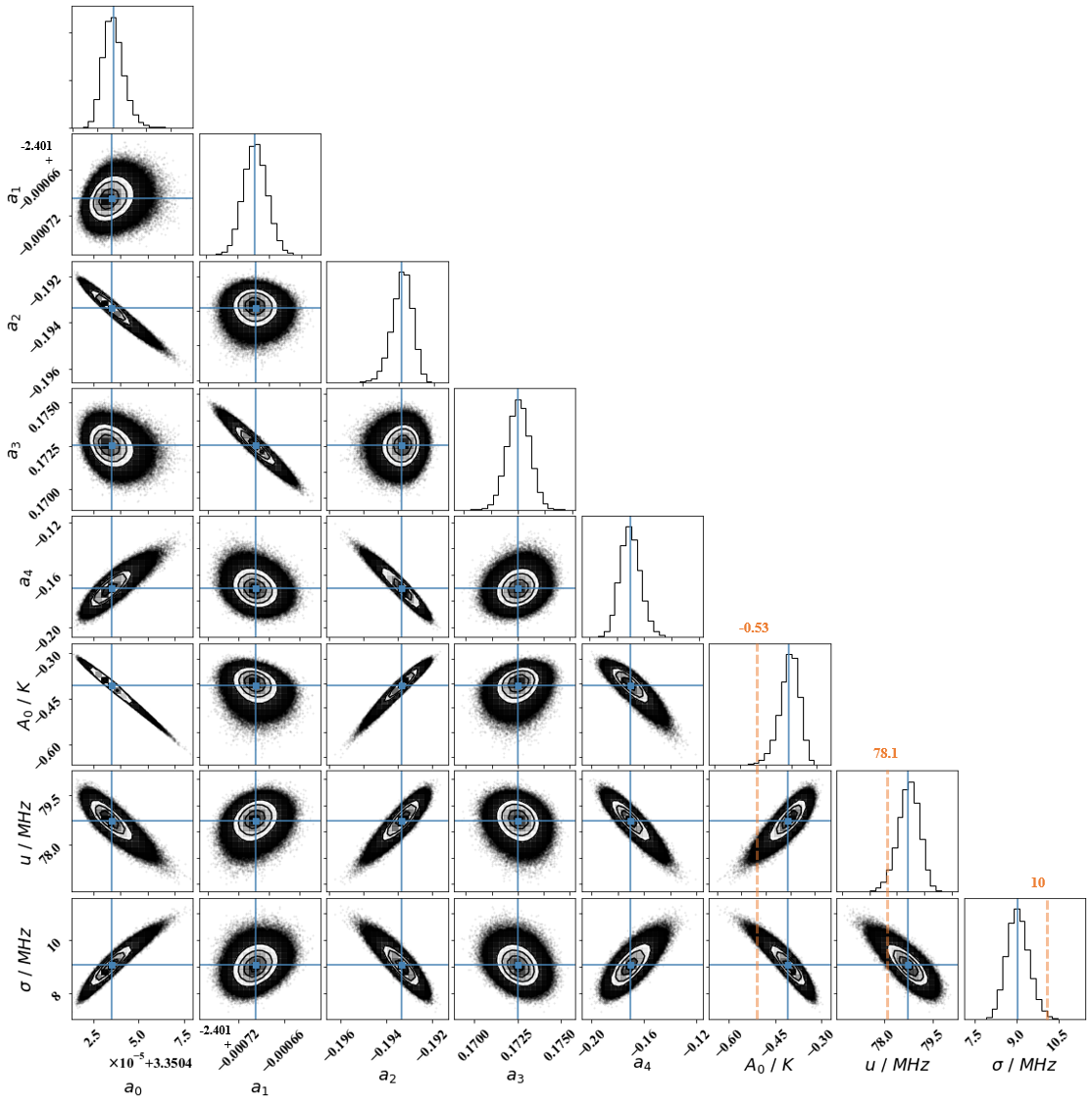}}
    \caption{ \label{fig:corner_plot2} \rev{
    The distribution of model parameters from the MCMC analysis, showcasing the outcomes of the logarithmic polynomial with a Gaussian signal model. 
    As depicted, the foreground coefficients of odd polynomial orders tend to be independent from those of even orders. However, within the subsets of even and odd coefficients, degeneracies persist. 
    Moreover, when compared to  \ref{fig:corner_plot1}, the signal parameters exhibit a greater degeneracy and non-Gaussianity. Notably, both the amplitude $A_0$ and trough location $u$ exhibit longer tails in their distributions leading to an enlarged error band around the best-fit signal. The true values of foreground parameters are indicated by an orange dashed line.} 
    }
    \end{figure*}

\bibliographystyle{aasjournal}
\bibliography{main}




\label{lastpage}
\end{document}